\documentclass[
  12pt,
  DIV=12,
  headings=standardclasses,
  headings=big
]{scrartcl}
\usepackage[utf8]{inputenc}

\usepackage[colorlinks=true,citecolor=gray]{hyperref}
\usepackage[backend=biber,sorting=none,sortcites,url=false,maxbibnames=5,backref=false,style=numeric-comp]{biblatex}
\usepackage{newtxtext,newtxmath,microtype}  
\usepackage{amsmath,amsfonts}
\usepackage{booktabs}
\usepackage{pgffor}

\title{A scalable and real-time neural decoder for topological quantum codes}%

\author{
\large
Andrew W.~Senior$^{1\dagger}$\footnote{Corresponding authors},
Thomas Edlich$^{1\dagger}$,
Francisco J.H.~Heras$^{1}$\footnote{Equal contribution},\\\large
Lei M.~Zhang$^{1}$,
Oscar Higgott$^{2}$,
James S.~Spencer$^{1}$,
Taylor Applebaum$^{1}$,\\\large
Sam Blackwell$^{1}$,
Justin Ledford$^{2}$,
Akvilė Žemgulytė$^{1}$,
Augustin Žídek$^{1}$,\\\large
Noah Shutty$^{2}$,
Andrew Cowie$^{1}$,
Yin Li$^{1}$,
George Holland$^{1}$,
Peter Brooks$^{2}$,\\\large
Charlie Beattie$^{1}$,
Michael Newman$^{2}$,
Alex Davies$^{1}$,
Cody Jones$^{2}$,\\\large
Sergio Boixo$^{2}$,
Hartmut Neven$^{2}$,
Pushmeet Kohli$^{1}$,
Johannes Bausch$^{1\dagger\ast}$
\\
\hspace{-1.4cm}%
\normalsize{$^{1}$Google DeepMind \& $^{2}$Google Quantum AI}
}

\date{}

\usepackage{fancyvrb}

\usepackage{graphicx}
\graphicspath{{figures/}}
\usepackage{makecell}

\usepackage[section]{placeins}

\input{braket}
\usepackage{amsfonts}
\usepackage[capitalise]{cleveref}
\usepackage{booktabs}
\usepackage{upgreek}
\usepackage[format=plain]{caption}
\usepackage{subcaption}
\begin{document}
\baselineskip20pt
\maketitle 
\thispagestyle{empty}
\enlargethispage{2cm}
\begin{refsegment}
\defbibfilter{notother}{not segment=\therefsegment}
\begin{abstract}

Fault-tolerant quantum computing will require error rates far below those achievable with physical qubits. Quantum error correction (QEC) bridges this gap, but depends on decoders being simultaneously fast, accurate, and scalable. This combination of requirements remains unmet by a machine-learning decoder, nor by any decoder for promising resource-efficient codes such as the color code.
Here we introduce AlphaQubit 2, a neural-network decoder that achieves near-optimal logical error rates for both surface and color codes at scale under realistic noise.
For the color code, it is orders of magnitude faster than other high-accuracy decoders. 
We demonstrate real-time decoding faster than $1\mathrm{\mu s}$ per cycle on commercial accelerators: for the surface code to distance 11, with better accuracy than leading real-time decoders; and the first real-time decoding of the color code to distance 9. 
These results support the practical application of a wider class of promising QEC codes, and establish a credible path towards high-accuracy, real-time neural decoding at the scales required for fault-tolerant quantum computation.
\end{abstract}

\clearpage
\baselineskip15pt
\setcounter{tocdepth}{2}
\tableofcontents
\baselineskip20pt

% MAIN TEXT
\clearpage
\section{Introduction}

Quantum computation promises significant advantages over classical computers, with potential applications in fields such as physics simulation \cite{Lloyd1996}, cryptography \cite{pirandola2020advances}, and optimization \cite{Farhi2001}. Realising this potential, however, depends on reliably executing complex algorithms on quantum hardware that is intrinsically noisy and error-prone.

The established route to reliable quantum computation is quantum error correction (QEC), which protects information by encoding a single logical qubit across multiple physical qubits \cite{Shor1995}. Below a critical physical error rate threshold, increasing this redundancy exponentially suppresses the logical error rate. Nevertheless, the magnitude of the challenge is substantial: fault-tolerant computations, such as the factorization of a 2,048-bit number, will require logical error rates below $10^{-10}$ per logical operation \cite{Campbell2022-pd,Kivlichan2019-az}, seven orders of magnitude below today's typical superconducting physical error rates \cite{Willow}.

Experimental progress towards fault-tolerance is accelerating, with recent demonstrations across multiple hardware platforms confirming the foundational principle of QEC. These efforts have primarily focused on two leading planar codes: the surface code, known to have the highest error threshold of all planar codes \cite{Fowler2012}; and colour codes, attractive for their more efficient logical operations \cite{Bombin2007-np} and lower error correction overhead when implementing a universal gate set \cite{bombin2006topological,Gidney2024-hi}. Recent experiments have demonstrated the suppression of logical errors with increasing code size in superconducting qubits (both in surface \cite{milestone2, Willow} and colour \cite{Lacroix2025} codes) and neutral atoms \cite{bluvstein2025architectural}. The practical viability of any code, however, depends critically on the classical processing component of error correction: the decoder.

QEC codes, including the surface and colour codes, produce a sequence of parity checks (the error syndrome), which the decoder must interpret in real time to decide whether the logical qubit needs correcting. To be viable, a decoder must simultaneously satisfy two stringent requirements. The first requirement is high decoding accuracy, even during logical operations \cite{Lacroix2025, Zhou2025}. For a given physical error rate, achieving higher accuracy requires scaling to larger code distances involving more physical qubits. The second requirement is a decoding speed faster than the hardware’s clock cycle (e.g.\ $\sim1~\mathrm{\upmu s}$ for superconducting devices \cite{Willow}, $\sim1~\mathrm{ms}$ for neutral atoms \cite{Bluvstein2023}) lest an exponential backlog render error correction infeasible \cite{terhal2015quantum}. While machine learning (ML) has emerged as a promising paradigm \cite{Bausch2024, NeuralDecoders, lange2023datadriven, Zhou2025, GenerativeDecoding, liu2025decoding, bluvstein2025architectural, hu2025efficient,zhang2025learning}, no ML-based decoder has yet met these combined demands, particularly for the promising but challenging-to-decode colour code \cite{gidney2023chromobius}.

Here we introduce AlphaQubit 2 (AQ2), a neural network decoder that makes substantial progress towards fulfilling both accuracy and speed requirements. AQ2 achieves near-optimal accuracy at large scales, reaching logical error rates below $10^{-10}$ per cycle for the surface and colour codes. For the colour code, it is orders of magnitude faster than available implementations of highly accurate decoders. Furthermore, we show that a real-time variant of AQ2 can decode both the surface code and colour code (up to 241 and 181 physical qubits respectively) in under $1~\mathrm{\upmu s}$ per cycle on commercial hardware, with little loss in accuracy. This combination of large-scale accuracy and real-time throughput up to intermediate code distances allows us to outline a viable path to real-time decoding of superconducting qubits at sufficient speed and scale to enable fault-tolerant quantum computation.

\section{Decoding surface and colour codes}

\begin{figure}[t]
\centering
\hspace{-3mm}
\includegraphics[trim={0 0 0 4mm},clip,height=8cm]{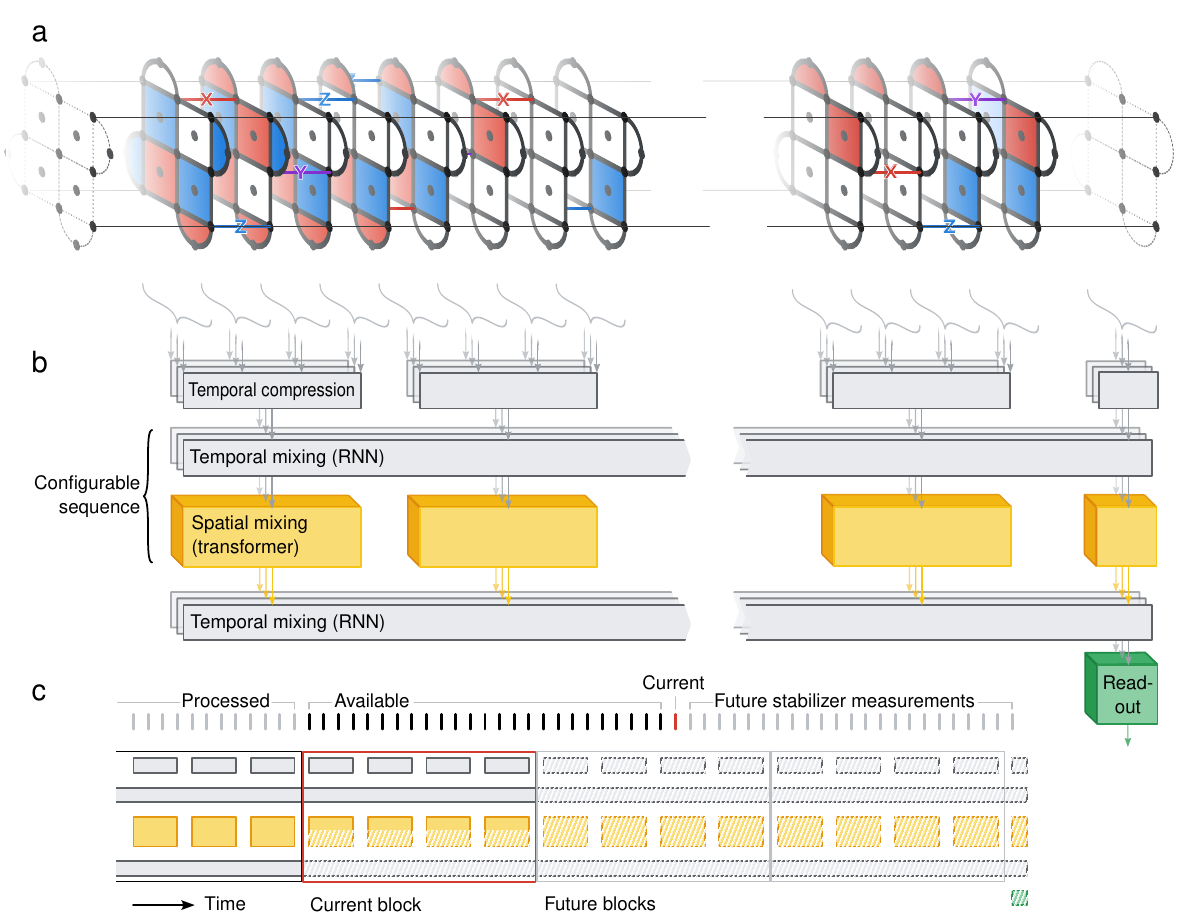}
\hfill
\includegraphics[trim={0 0 3mm 0},clip,height=8.05cm]{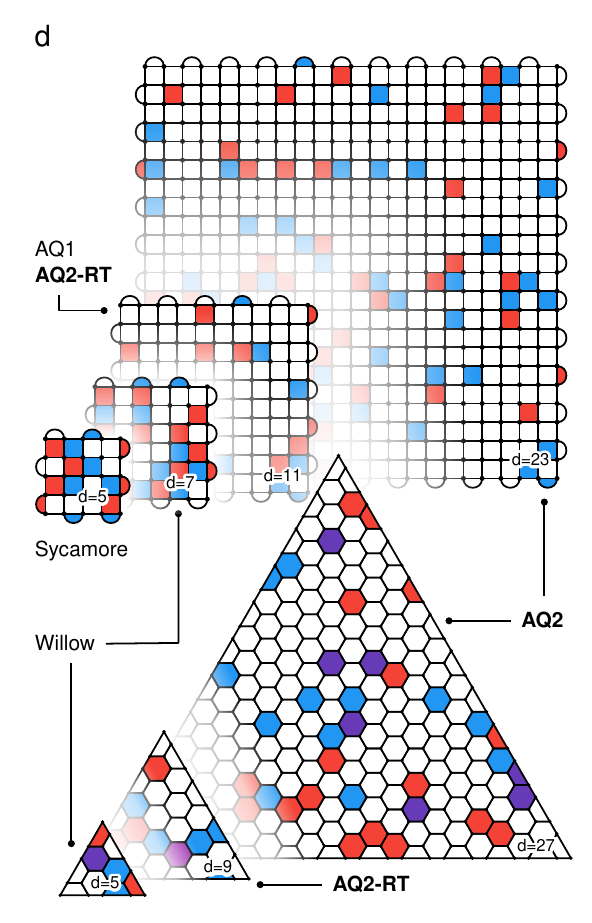}
\caption{
\textbf{Decoding memory experiments with AlphaQubit 2 (AQ2).}
\textbf{a,} In a surface code memory experiment, a logical qubit is initialized; repeated stabilizer checks are performed; and then the logical qubit state is measured. During the experiment all qubits and operations are subject to errors (here symbolically shown as bit (X), phase (Z), and combined bit and phase flips (Y) acting on individual data qubits between time steps). These errors affect the stabilizer parity checks (X check failure in red, Z check failure in blue).
\textbf{b,} Overview of the AQ2 architecture. Information flows from top to bottom and left-to-right. Consecutive checks for each stabilizer are temporally compressed in groups, then processed by alternating temporal and spatial mixing layers. The output from the last time step is fed to a readout network which makes a prediction of the logical error. 
\textbf{c,} Spatial and temporal mixing updates can be done in a streaming fashion: different temporal windows can be computed as stabilizer measurements are available, without waiting for the experiment to end.
\textbf{d,} Logical qubit patches for the surface and colour codes (distance 23 and 27, resp.) as decoded by AQ2. Superimposed are the largest surface code sizes previously implemented on Sycamore (distance 5 \cite{milestone2}); Willow (distance 7 \cite{Willow}) and decoded by AlphaQubit 1 (distance 11 \cite{Bausch2024}) and the distance-5 colour code implemented on Willow \cite{Lacroix2025}; the distance 11 surface code and distance 9 colour code can be decoded in real-time by AQ2-RT, a real-time variant of AQ2.}
\label{fig:sc-and-memory-experiment}
\end{figure}

The surface code has been the primary focus of experimental QEC on superconducting hardware (Fig.~\ref{fig:sc-and-memory-experiment}a,d). It encodes a logical qubit on a grid of $d \times d$ data qubits, and uses an interspersed set of $d^2-1$ measure qubits to periodically execute X and Z stabilizer parity checks on adjacent data qubits. A detection event occurs when stabilizer checks in two consecutive cycles of measurements disagree (Fig.~\ref{fig:sc-and-memory-experiment}a). The logical state of the surface code is determined by a pair of observables, $X_L$ and $Z_L$, which anti-commute with each other and commute with all stabilizers. The side length, $d$, of the grid is the same as the code distance, i.e. the minimum number of physical qubit errors required for an undetectable logical error.

Decoding the surface code is a mature field, dominated by the Minimum Weight Perfect Matching algorithm (MWPM) and its variants \cite{edmonds1965paths,fowler2013optimal} (e.g.\ PyMatching \cite{pymatching, higgott2025sparse}). While highly efficient, MWPM provides only an approximate solution.
Recently, augmented variants of MWPM have demonstrated higher accuracy by ensembling matching solutions (Libra \cite{Libra} and Harmony \cite{Harmony}) or by generalizing MWPM to hypergraphs (Minimum-Weight Parity Factor \cite{wu2025minimum}). Tesseract \cite{Beni2025} can achieve near-optimal accuracy by using A*-search to find the most likely error source, at considerable computational cost.
This highlights a persistent trade-off between decoding speed and accuracy \cite{deMarti-iOlius2024-zy, wu2025minimum}.

Colour codes are an alternative that may require fewer physical qubits and facilitate implementation of logical operations \cite{Bombin2007-np,bombin2006topological,Gidney2024-hi}. 
In this paper we use both the Bell-flagged \cite{chamberland2018flag, Baireuther2019} and the superdense \cite{gidney2023chromobius} variants of the triangular colour code, where data qubits are arranged on the vertices of a honeycomb lattice (Fig.~\ref{fig:sc-and-memory-experiment}d). Each hexagonal cell defines one X and one Z stabilizer. In the superdense variant, both are measured synchronously with the aid of two measure qubits. In the Bell-flagged variant, one of two measure qubits per cell acts as a ``flag'', and the stabilizers are read in an alternating fashion; one full error correction cycle thus takes about twice as long as for the superdense variant.

Developing effective decoders for colour codes has been a significant challenge, which has likely played a role in hampering their experimental adoption \cite{Lacroix2025}. Existing decoders again illustrate a trade-off between speed and accuracy. While Tesseract \cite{Beni2025} can be applied to colour codes, it is very computationally intensive. In contrast, Chromobius \cite{gidney2023chromobius} has been specifically designed to work with the colour code and is fast but has considerably lower accuracy. More recent approaches fall between these two extremes \cite{Koutsioumpas2025,Lee2024}. To date, the lack of a decoder that is simultaneously fast, accurate, and scalable has been a significant barrier to leveraging the colour code’s potential advantages.

\begin{figure}[t]
\centering
\includegraphics[width=\textwidth]{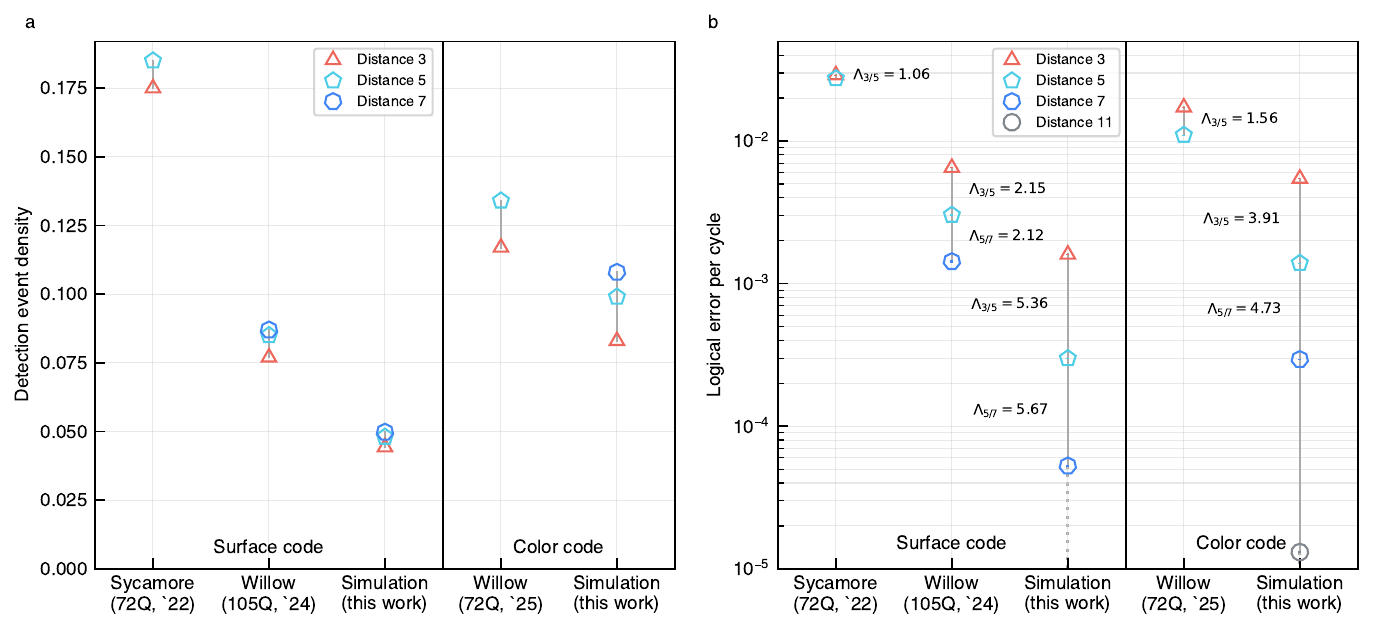}
\caption{
\textbf{Progress in superconducting hardware. a,} Detection event fraction for surface and colour codes of different distances implemented in two generations of superconducting quantum computing (Sycamore \cite{arute2019quantum} and Willow \cite{Willow,Lacroix2025}), compared to the simulated circuit depolarizing noise model (SI1000 \cite{Gidney2021honeycomb} at $p=0.15\%$) chosen for the experiments in this paper, anticipating future hardware improvements. \textbf{b,} The corresponding progress in logical error rate for surface and colour codes, when decoded using AlphaQubit 1 \cite{Bausch2024,Willow,Lacroix2025} (Sycamore and Willow) and AlphaQubit 2 (this work).}
\label{fig:noise-and-lambda}
\end{figure}

\section{High accuracy decoding}
\subsection{Neural network decoding}

Neural networks are a promising approach for QEC decoding. Diverse strategies, from generative \cite{GenerativeDecoding, liu2025decoding} to discriminative \cite{Bausch2024} models, have been successfully applied to decoding memory experiments in surface, colour and qLDPC codes \cite{lange2023datadriven, Lacroix2025}, as well as logical gate operations \cite{Lacroix2025,NeuralDecoders,Zhou2025}. While these studies  establish the potential of machine learning, they have not yet delivered a decoder that simultaneously achieves the high accuracy and speed required for fault-tolerant quantum computation.

Achieving the logical error rates of $10^{-10}$ or lower per cycle required for practical fault tolerant quantum algorithms \cite{Campbell2022-pd,Kivlichan2019-az,gidney2025factor2048bitrsa} necessitates scaling QEC codes to distances beyond the reach of near-term hardware. Rigorous assessment of decoder performance at these scales therefore depends on large-scale simulation. We generate training, evaluation and test data using Stim \cite{gidney2021stim}, employing the superconducting-inspired circuit-level depolarizing noise model SI1000 \cite{Gidney2021honeycomb}. We choose a physical noise level ($0.15\%$) corresponding to an anticipated improvement in hardware performance comparable to the progression from Sycamore to Willow quantum processors. In the surface code, this leads to a detection event density of $\sim 5\%$ (Fig.~\ref{fig:noise-and-lambda}a) and an error suppression factor of $\gtrapprox 5$ (Fig.~\ref{fig:noise-and-lambda}b).

\subsection{A scalable spatiotemporal architecture}

AQ2 processes the error syndrome of a logical qubit in a streaming fashion and predicts the logical observable. We designed its architecture (Fig.~\ref{fig:sc-and-memory-experiment}b) to be applicable to experiments of arbitrary code distance and duration, and to leverage accelerator parallel processing both during training and evaluation.

The network is composed of interleaved temporal and spatial layers which compute internal representations for each stabilizer at each error correction cycle. Recurrent neural-network layers implement temporal updates by propagating information forward in time for each stabilizer independently (but sharing parameters), integrating new measurement data with the stabilizer's previous state. 
Spatial updates are performed by attention-based transformer layers that allow stabilizer representations to exchange information with each other across the entire code patch at a single instant in time. 
After the final cycle, a readout network aggregates the final state vectors from all stabilizers to predict the probability of a logical error.

To achieve higher speeds both during training and evaluation, instead of performing a temporal update at each measurement cycle, we combine groups of (typically 3 to 6) consecutive measurement cycles, using a learned temporal compression which results in no loss in accuracy.  Processing can be carried out on a block of such groups, in which the computation for all the groups can be carried out in parallel for each spatial transformer layer. Since the architecture propagates information causally, with only the recurrent state vectors carrying information about the syndrome history forward, the computations can be performed in a streaming fashion across blocks, while new stabilizer measurements are received from the quantum computer (Fig~\ref{fig:sc-and-memory-experiment}c).

We train AQ2 following a novel training regime designed for scalability and accuracy; using gradient descent on many millions of examples generated by Stim \cite{gidney2021stim} across a wide range of code distances, error rates, and experiment durations.
To enable robust and shorter training, we employ a curriculum learning strategy, where the model is first trained predominantly on easier examples (smaller code distances, short experiments) before progressing to more challenging examples (larger code distances). For the highest code distances, we further fine-tune the model exclusively on examples for a single code distance and ensemble two or more variants of the model. Furthermore, we introduce two key techniques to stabilize and accelerate learning: a novel auxiliary loss that tasks the network with predicting an idealized, noise-free readout of the logical observable, and a random input dropout method. Details of the network architecture and training process are described in Methods.

\subsection{Decoding at scale}

\begin{figure}[t]
\includegraphics[width=\textwidth]{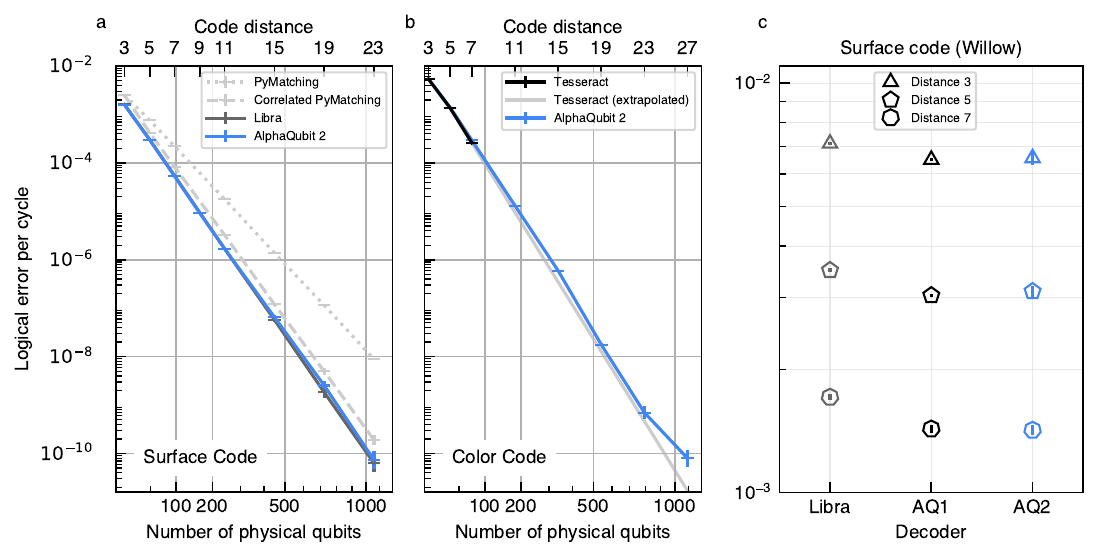}
\caption{
\textbf{AlphaQubit 2 (AQ2) accuracy on simulated and experimental data.} Accuracy at scale: Logical error per cycle against number of physical qubits / code distance for the surface code \textbf{a,} comparing AQ2 with the Libra decoder and PyMatching and for the Bell-flagged colour code \textbf{b,}, comparing AQ2 with extrapolated Tesseract error rates. Measured on up to $2.5 \times 10^{10}$ shots of 120 cycles from the SI1000 noise model (details in Methods). 
\textbf{c,} Experimental data: logical error per cycle at code distances 3, 5 and 7 for AQ2 on the Willow experimental data \cite{Willow} compared to the most accurate decoders from the original paper (AlphaQubit 1 \& Libra).  
Error bars are 95\% confidence intervals in all figures.}
\label{fig:accuracy}
\end{figure}

AQ2 achieves near-optimal accuracy on the surface code up to code distance 23. On simulated 120-cycle memory experiments with 0.15\% circuit-level noise, it achieves a logical error per cycle of $7.3\times10^{-11}$  (Fig.~\ref{fig:accuracy}a).
This performance is close to that of the near-optimal, but slower, Libra decoder and substantially outperforms both correlated and uncorrelated PyMatching decoders \cite{pymatching, higgott2025sparse}.

Furthermore, AQ2 achieves high-fidelity decoding of the Bell-flagged colour code at large scale (Fig.~\ref{fig:accuracy}b).
There is no fast, high-accuracy decoder for the colour code. We compare AQ2 performance to the near-optimal Tesseract \cite{Beni2025} decoder at small distances ($d \le 7$). Beyond this distance Tesseract is too slow to evaluate so we extrapolate linearly from distance 5 and 7 to estimate an ideal decoding accuracy.
Our results show that, for the Bell-flagged code, AQ2 tracks this idealized performance closely up to distance 23. At distance 27, an ensemble of 3 models reaches a logical error per cycle of $8.0\times 10^{-11}$, but does not match the idealized trend in this regime where training AQ2 is more difficult and expensive (see Methods for details). High-accuracy decoding is also achieved for the superdense colour code (Fig.~\ref{fig:realtime}b).

\subsection{Decoding the Willow experiment}

To validate our decoder with data from a physical quantum processor, we applied AQ2 to experimental data from the 105-qubit Willow chip \cite{Willow}, which implements surface codes of distance 3, 5, and 7.
We find that our new decoder achieves a logical error rate comparable to that of our previous, slower architecture, AlphaQubit \cite{Bausch2024} (AQ1), and better than Libra on the same experimental dataset (Fig.~\ref{fig:accuracy}c). This result was obtained using the three-stage training protocol detailed in our previous work \cite{Willow}: pretraining on generic simulated noise (SI1000); fine-tuning on simulated samples from a Detector Error Model (DEM) fitted to experimental data; and further fine-tuning on a subset of the experimental samples from the Willow hardware itself.

\section{Real-time decoding}

A primary requirement for practical QEC is real-time decoding, where measurement data must be processed faster than it is generated to avoid incurring an exponential backlog \cite{terhal2015quantum}. For superconducting hardware this imposes a demanding average throughput rate of around $1~\mathrm{\upmu s}$ per cycle. 

\begin{figure}[t]
\centering
\includegraphics[trim={2mm 2mm 2mm 0},clip,width=\textwidth]{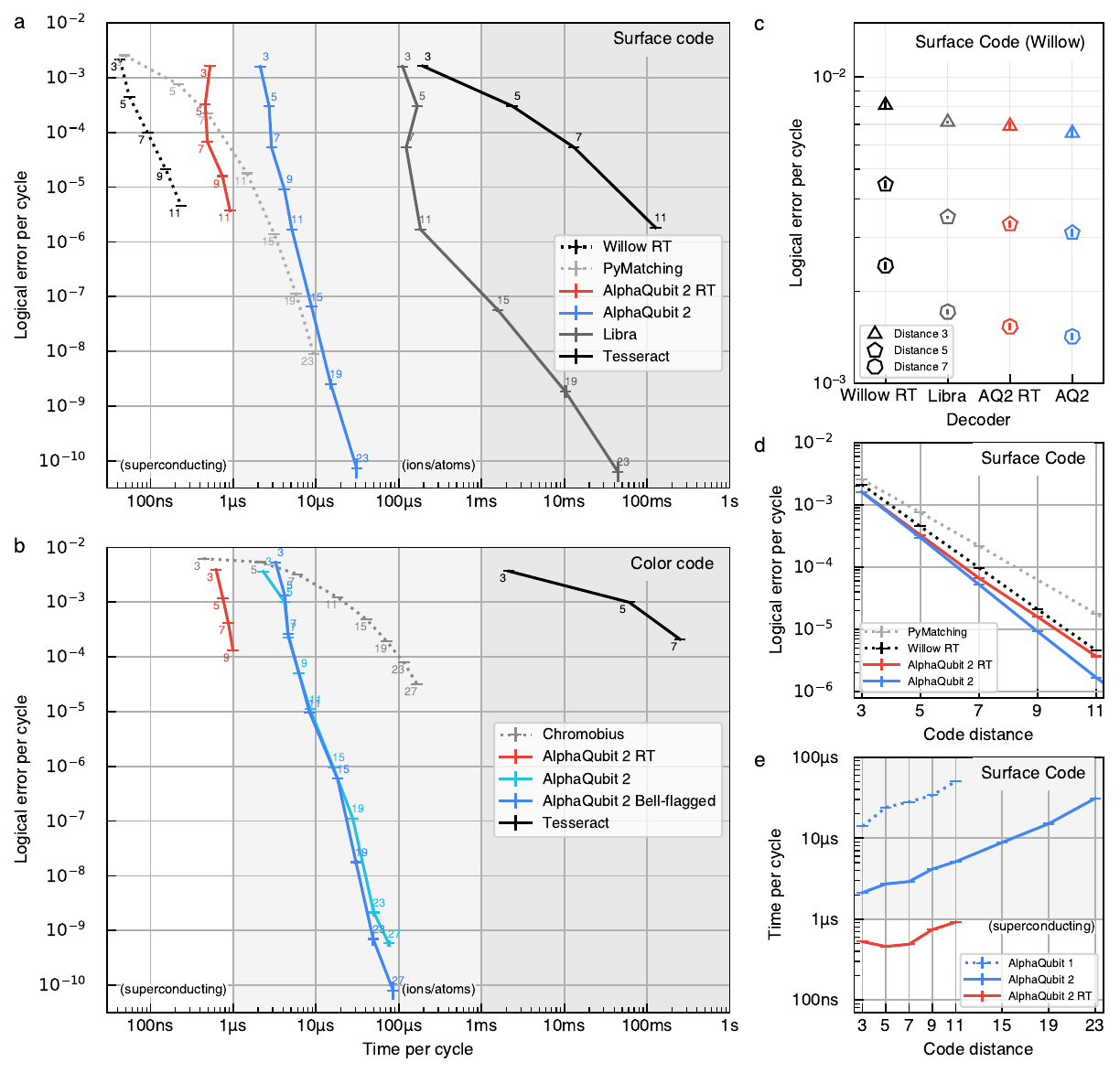}
\vspace*{-\baselineskip}\caption{
\textbf{Accurate, high-throughput decoding of surface and colour codes with AlphaQubit 2 (AQ2). a,} Logical error per cycle vs.~throughput (average time per cycle) when decoding surface codes across code distances with real-time AQ2 RT (red), full AQ2 (blue), and other decoders (gray) on SI1000 data (0.15\% noise). Shading indicates the $1~\mathrm{\upmu s}$ (resp. $1~\mathrm{ms}$) target speeds for superconducting (resp. other hardware substrates) with surface codes. 
\textbf{b,} 
As (\textbf{a}) but for the superdense colour code, plus AQ2 on the Bell-flagged colour code (dark blue). 
\textbf{c,} Logical error per cycle at distance 3, 5 and 7 for AQ2 RT on the Willow experimental data \cite{Willow} compared to the real-time matching decoder, full AQ2 and original paper decoders (AQ1 \& Libra).
\textbf{d,} Logical error per cycle vs.\ code distance on SI1000 data (0.15\% noise) for the real-time and full AQ2 compared to Tesseract and PyMatching.%
\textbf{e,}  Throughput vs.\ code distance for three AlphaQubit versions. AQ2 is timed on Trillium TPU, AQ1 on the earlier TPU v5e.
}
\label{fig:realtime}
\end{figure}

AQ2 as described above, in both the surface and the colour code, is more accurate than any faster decoder we compared it against, \textit{i.e.} it sits on the Pareto front of the accuracy-throughput trade-off (Fig.~\ref{fig:realtime}a,b). For code distances up to 27, AQ2 has a cycle time below $100~\upmu\mathrm{s}$: fast enough to decode neutral atom or trapped ion circuits (around $1~\mathrm{ms}$\cite{Bluvstein2023,ryan2021realization}), but too slow to decode superconducting circuits in real time.

To enable decoding on superconducting qubits, we have developed a real-time variant of AQ2 (AQ2-RT). This is a compact version of AQ2 with fewer layers, smaller representations and faster recurrence (see Methods). AQ2-RT achieves this stringent real-time throughput for the surface code up to distance 11 (Fig.~\ref{fig:realtime}a,d) and for the superdense colour code up to distance 9 (Fig.~\ref{fig:realtime}b), using commercially-available machine learning accelerators (Trillium TPUs \cite{Trillium}), without the need for custom hardware such as ASICs or FPGAs. 

The throughput improvements of AQ2-RT are substantial (Fig.~\ref{fig:realtime}). For the surface code, while the full AQ2 is already 9.6 times faster at distance 11 than AQ1, the real-time variant provides a further 6 times speed-up (Fig.~\ref{fig:realtime}a,d). 
Both AQ2 and AQ2-RT are streaming decoders whose computational costs do not vary with the errors observed, and whose throughputs can be maintained for longer experiments without any increase in latency, which remains below $500~\mathrm{\upmu s}$ (see Methods).

For the surface code, the speed increase of AQ2-RT entails a modest trade-off in accuracy. The real-time decoder's logical error rate is slightly higher than that of the full-accuracy model (Fig.~\ref{fig:realtime}a,c), particularly at distance 11 ($3.73\times10^{-6}$ vs $1.89\times10^{-6}$). At that code distance, the real-time model remains more accurate than other decoders which can achieve this throughput (a real-time matching decoder \cite{Willow} and PyMatching for lower code distances). 

The accuracy drop of the real-time model is also small on experimental data. To assess this we trained AQ2-RT on the Willow data, with the same pretraining and fine-tuning protocol used above (Fig.~\ref{fig:realtime}C). AQ2-RT accuracy is close to that of the full AQ2, and is more accurate than both the original real-time decoder used on this dataset (Willow RT; Fig.~\ref{fig:realtime}C) and the Libra decoder. (See Methods for details of the models and timing.)

For the colour code, the AQ2-RT model is real-time up to distance 9 (Fig.~\ref{fig:realtime}B), but does incur more of an accuracy drop compared to full AQ2. Note that we adopt a superdense variant whose planar readout facilitates superconducting circuit implementation \cite{Lacroix2025} as well as decoding with Chromobius \cite{gidney2023chromobius}, the fastest colour-code decoder available, which has higher error rates and achieves real-time  throughput only at distance 3. Full AQ2 achieves the best accuracy with the Bell-flagged variant at large code distances, and with the superdense variant for low code distances (Fig.~\ref{fig:realtime}B).

\section{Robustness}

\begin{figure}[t]
\centering
\includegraphics[trim={0 0 0 0},clip,width=\textwidth]{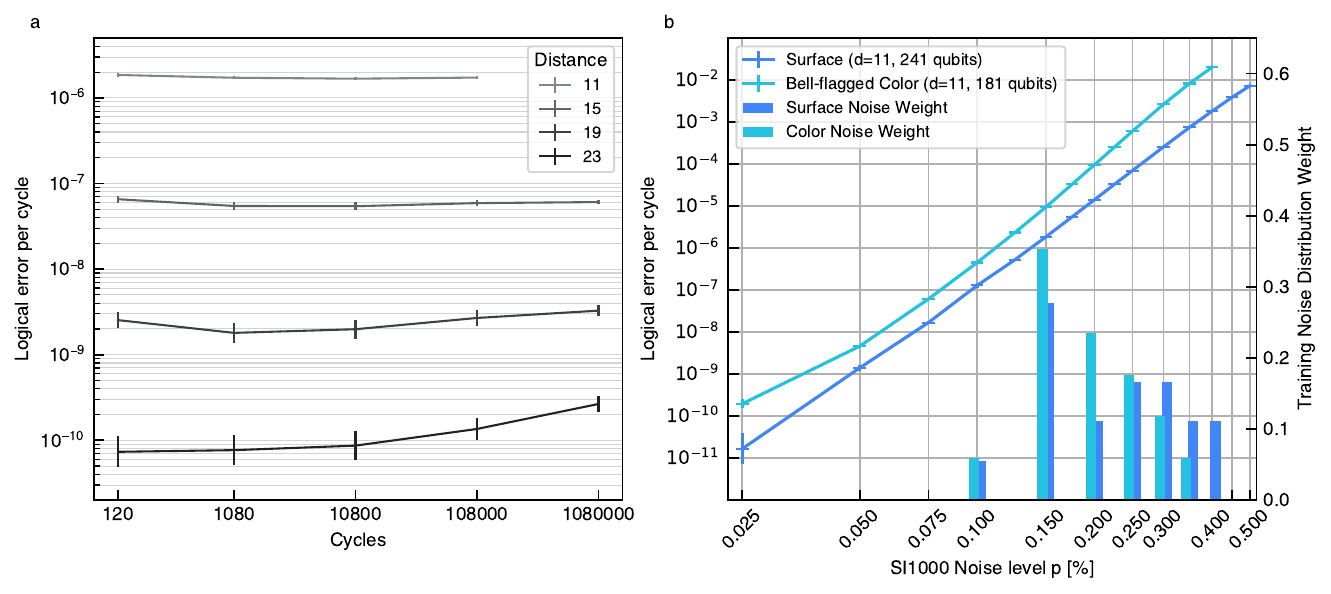}
\caption{
\textbf{Generalisation beyond the training distribution. a,} 
Decoding longer experiments. Decoding performance against number of cycles for AlphaQubit 2 measured on the surface code at code distances 11--23, with SI1000 $p=0.15\%$ noise. The total number of shots used for evaluation increases with code distance (see Methods). \textbf{b,} logical error per cycle against noise level for distance-11 surface- and Bell-flagged colour-code models (241 and 181 qubits). The bar plot represents the noise level distributions used during training.
}
\label{fig:longrounds}
\end{figure}

A critical requirement for any practical decoder is temporal stability: it must perform reliably over millions of QEC cycles, far longer than is feasible for training.
AQ2's time-invariant, recurrent architecture is explicitly designed for this scenario, enabling it to operate in a streaming fashion with a fixed computational cost per cycle and constant memory requirement.
To test this generalization capability, we took a model trained on experiments of maximum length 168 cycles and evaluated it on experiments up to one million cycles. AQ2 maintains a stable logical error per cycle, albeit with a small degradation in performance at distances 19 and 23 (Fig.~\ref{fig:longrounds}a). 

AQ2 is also robust to changes in sample noise strength, without retraining or requiring the noise level as an explicit input. For instance, the decoders for both colour and surface codes perform well across at least a sixteen-fold variation in circuit noise strength, including values above and below the range of circuit noise used for training (See Fig.~\ref{fig:longrounds}b for distance 11).

\section{Discussion and conclusion}

In this work, we introduced AQ2, a neural-network decoder that addresses the dual requirements of high accuracy and real-time speed for quantum error correction.
We have shown that this scalable architecture achieves near-optimal logical error rates for both the surface code and colour codes up to distance 23, matching or exceeding the accuracy of the best available decoders, and its time-invariant design was shown to generalize to million-cycle experiments, far exceeding the longest training example.
Crucially, we demonstrated that a compact version of AQ2 meets the stringent $1~\mathrm{\upmu s}$ cycle time required for superconducting hardware for surface and colour codes up to distances 11 and 9 respectively, using commercially available accelerators, and with little accuracy loss.
The decoder's performance was further validated on experimental data from a real-world quantum device, for which AQ2 (full and RT) had lower error rates than our strongest baseline. This demonstrates the advantage of learning from data when decoding in the presence of correlations not captured by local Pauli error models.

While our results represent significant progress, several challenges remain.
In terms of accuracy, we observed a deviation from ideal scaling at the highest distances, where training to convergence becomes more challenging, particularly for the colour code at distance 27 and in million-cycle surface code experiments, indicating that further work on training methodologies for these regimes is needed.
Reaching the logical error rates of $10^{-12}$ and below that are needed for large-scale quantum applications is a further hurdle for which even reliable evaluation is a challenge, which will require increased computational resources or more sophisticated statistical estimation \cite{mayer2025rareeventsimulationquantum}.
Although we achieved real-time throughput for distances up to 11, extending this capability to the larger distances required for full fault-tolerance will be the next major milestone. We anticipate that this can be achieved through a combination of further architecture innovation, algorithmic co-design, the use of low-precision arithmetic, and possibly windowing \cite{zhang2025learning} or implementation on specialized hardware (FPGAs, ASICs) as large-scale quantum devices become available (see Methods for an extended discussion). We have not yet addressed reducing the latency between the final measurement and the decoding result, whereas for many quantum computing applications it is critical to keep this to a minimum.  Finally, while we have previously demonstrated the ability to decode logical operations (for the colour code \cite{Lacroix2025}), integrating this capability at scale will be necessary for a complete fault-tolerant quantum computing stack. 

In conclusion, AQ2 represents a substantial advance in quantum error decoding and demonstrates that neural decoders can satisfy the core requirements for fault-tolerant quantum computing. By providing a viable path to real-time neural decoding, our work addresses a critical bottleneck for leading error correction codes, while the model's code-agnostic architecture opens the potential for discovering more efficient codes and protocols, accelerating the timeline towards practical quantum computers.

\clearpage
\printbibliography[segment=\therefsegment]
\section*{Acknowledgments}
We thank Aria Shahingohar, Borislav Kozlovskii, Craig Donner, Laleh Beni, Sebastian Bodenstein, Sebastian Nowozin, and Stig Petersen for their support and contributions to the project, and Alex Gaunt, Ryan Babbush and Vlad Sivak for further feedback on the paper. We thank Demis Hassabis for sponsoring the research and creating the environment that made this work possible.

\end{refsegment}

% APPENDIX
\clearpage
\appendix
\setcounter{figure}{0}
\setcounter{table}{0}
\renewcommand\thefigure{S\arabic{figure}}    
\renewcommand\thetable{S\arabic{table}}
\begin{refsegment}
\section{Methods}
\subsection{Network Architecture}\label{sec:network}
\subsubsection{Full AlphaQubit 2 architecture}

\begin{figure}[t]
\centering
\includegraphics[width=0.7\textwidth]{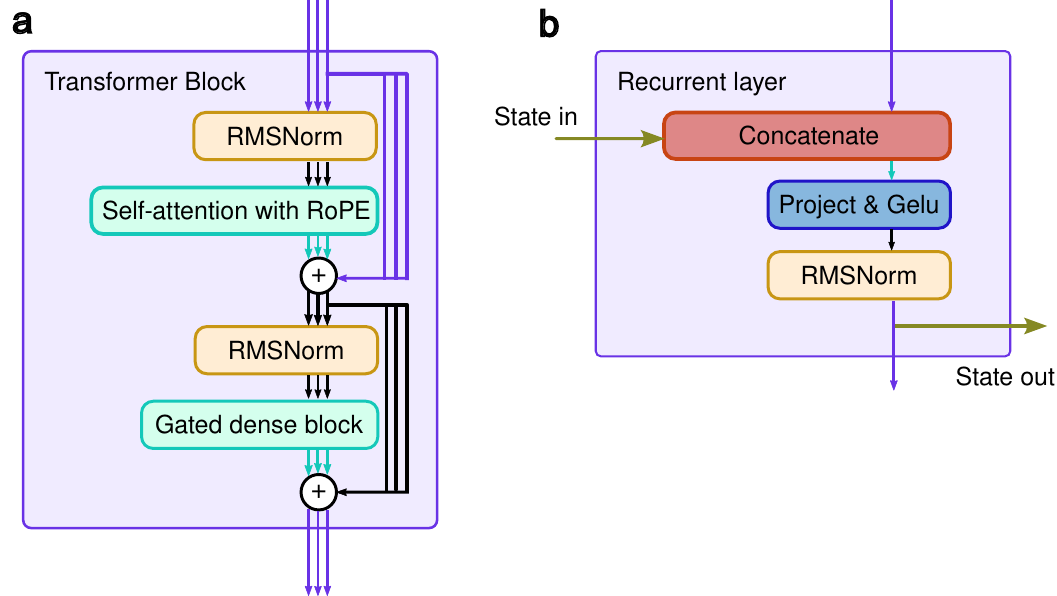}
\caption{
\textbf{Components of the AlphaQubit 2 architecture. a,} Spatial mixing transformer block operating on all stabilizer representations at one time step. \textbf{b,} Recurrent layer for each stabilizer.}
\label{fig:architecture_blocks}
\end{figure}

The AlphaQubit 2 (AQ2) architecture is a new architecture for predicting logical observables from stabilizer measurement events. It shares some attributes with AlphaQubit 1 (AQ1\cite{Bausch2024}) ---a per-stabilizer representation; recurrency; the use of transformers; and a readout network with pooling--- but is fundamentally different (\cref{fig:sc-and-memory-experiment}). In particular, instead of the transformer computation being inside the recurrent processing block, we now adopt a light-weight recurrent layer (\cref{fig:architecture_blocks}b) which we interleave with transformer layers (\cref{fig:architecture_blocks}a). This has advantages for inference speed, since now, at inference time, the transformer operations can be carried out in parallel across a number of cycles leading to more efficient use of the computing accelerator and thus greater throughput. For simplicity and uniformity of the architecture across code types, we use no convolutions and replace the readout network’s pooling from a data qubit representation with a simple mean pooling across stabilizers, with a separate embedding per-logical observable and cross-attention between the logical observable representations and the final per-stabilizer representation.  By training the network on the binary logical observable, the network outputs an estimate of the probability of the logical observable being 1.

Instead of processing one cycle in each time step, to improve inference and training speed, we combine multiple (6 for simulated surface code, 3 for simulated colour code, 5 for the Willow data) cycles of measurements, and run the network at a correspondingly lower rate. The final cycle of syndrome measurements in each experiment, derived from measuring the data qubits, is embedded by itself but processed with the same network.
The new architecture accumulates groups of cycles of measurements before temporally compressing them (by concatenating and projecting the vector embeddings)  and feeding the combined representation through a series of spatial or temporal network layers. The spatial layers are transformer layers that implement self-attention between the stabilizer representations (independently at each time step). The temporal layers are recurrent layers that integrate new information from a lower layer into a recurrent network state (independently for each stabilizer). At the end of an experiment, the final cycle of stabilizers is processed with the same network stack and the output from the final layer is supplied to the readout network to make a label prediction for each logical observable. 

For the full-sized AQ2, optimized for accuracy (AQ2-full), the recurrent layers (RNNs, Fig.~\ref{fig:architecture_blocks}b) combine the previous state and the layer’s inputs for the current cycle by concatenating them and projecting to the original dimension. The projection is followed by a Gelu activation\cite{hendrycks2016gaussian} and an RMSNorm layer\cite{zhang2019root}. The RNN output is used both to feed to the next layer and as the state for the next time step. Better training was observed by initializing the weights that project the state to a random orthogonal matrix. Each state is initialized with the zero vector at the start of an experiment. In the speed-optimised version of AQ2 (AQ2-RT) we used a faster, element-wise gated recurrence (See \ref{sec:full_vs_rt} for details).

The spatial mixing transformer layers (Fig.~\ref{fig:architecture_blocks}a) consist of an RMSNorm normalization followed by multiheaded self-attention between the stabilizer embeddings, with a parallel residual connection. A further normalization followed by a gated dense block computes a further update to the residualized activations.

For AQ2-full we use the following layer sequence, as shown in \cref{fig:one_block}:

\begin{Verbatim}[samepage=true]
RNN; RNN;
3 transformer layers; RNN;
3 transformer layers; RNN; 
3 transformer layers; RNN
\end{Verbatim}

\begin{figure}[t]
\centering
\includegraphics[width=\textwidth]{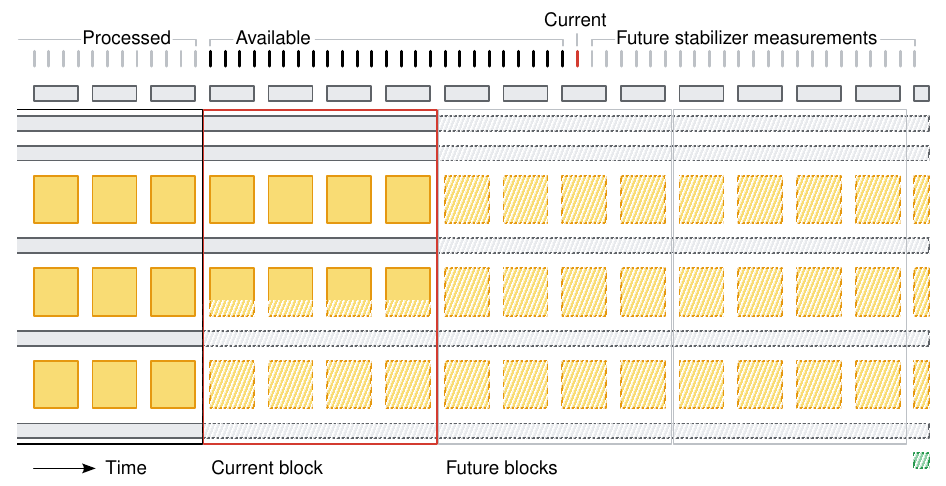}
\caption{
\textbf{Neural network layers for one time block of cycles.}}
\label{fig:one_block}
\end{figure}

The readout network combines the final per-stabilizer representation by mean-pooling. We replicate this representation for each logical observable being predicted (only one in inference) and add a learned embedding to each. The per-logical observable representations cross-attend to the final per-stabilizer representation.

After two cross-attention transformer layers, each logical-observable representation is processed by two residual dense layers before projecting to a single channel and logistic activation function.

\subsubsection{Embedding}
AQ2 uses a variety of embedding methods. 

Stabilizers are embedded with a linear embedding. For the surface code we add the embeddings of both measurement and events for each stabilizer, and for the colour code we add the embeddings of measurements and, for the Bell-flagged variant, the corresponding flags (See \ref{sec:unified}). We also add a normalized position encoding for each of $x$ and $y$, which linearly embeds the normalized coordinates of each qubit relative to the current code patch size. Thus,  for the surface code, regardless of code distance,  the left (right) edge always has an $x$ coordinate of -0.5 (0.5).

For spatial self-attention we use Rotational Position Encodings (RoPE\cite{rope}) of the spatial coordinates of the physical qubits. Half of the channels use the $x$ coordinate and half use the $y$ coordinate. For the colour code X \& Y basis measurements occur at the same physical location, so we add a basis embedding vector learned for each basis.
After pooling for the readout network, we add a separate learned embedding for each of the logical observables to disambiguate them for the cross-attention. 

\subsubsection{Ensembling}
In AQ1 we showed that even better accuracies can be obtained by combining the error estimates of multiple models with no time penalty, at the cost of additional hardware to parallelize the model evaluations. 
The same technique can be applied to AQ2. For simplicity, most results use a single model and we only use ensembling to achieve peak performance at the highest distances (\cref{tab:ensemble_hyperparameters}), and with the Willow experimental data, where we follow our previous protocol\cite{Willow} in which we ensemble 5 similar models trained with different random seeds.

For the surface code at distance 23, we ensemble a base model trained with code distances up to 23 with the same model further fine-tuned at distance 23. We note that the fine-tuned model achieved slightly higher accuracy at 120 cycles, but did not generalize as well to higher code distances. For the Bell-flagged colour code we ensembled fine-tuned variants of the base model: two at distances 19 and 23, and three at distance 27.  For the superdense colour code we ensembled the base model with three variants at distance 23 and 27. We would expect some improvement in all the results by ensembling where a single model was used, or increasing the size of the ensemble, but expect that most of the gains have been realised.
With distillation\cite{hinton2015distilling} it may be possible to train a single ``student'' model which achieves similar performance to the ensemble and thus avoid the additional computational burden.

\subsubsection{Principal architecture differences compared to AlphaQubit 1}
The principal differences between the AQ1 architecture and that used here for AQ2-full are as follows:
\begin{itemize}
    \item Adoption of the layer-based architecture with light-weight RNN layers (\cref{fig:architecture_blocks}b), rather than the processing all being carried out within a large RNN core.
    \item Temporal compression of consecutive frames before processing.
    \item No convolutions. 
    \item Use of cross-attention at the start of the readout network.
    \item Use of spatial RoPE.
\end{itemize}

\subsubsection{Architecture changes for Real-time AlphaQubit 2}
\label{sec:full_vs_rt}

\begin{figure}[t]
\centering
\includegraphics[height=8cm]{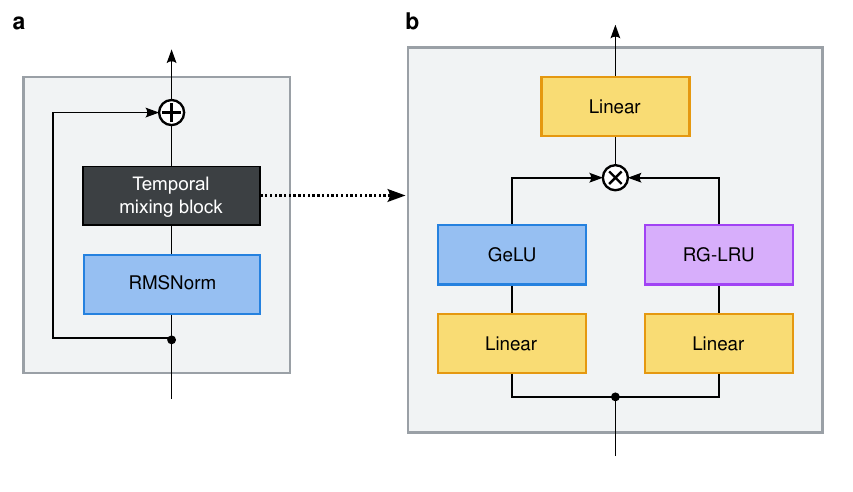}
\caption{
\textbf{Griffin RNN layers.} \textbf{a}, residual block. \textbf{b}, recurrent block.}
\label{fig:griffin}
\end{figure}

To achieve throughput of under $1\mathrm{\upmu s}$, we train smaller models (AQ2-RT) with the following variations: 
\begin{itemize}
    \item The layer sequence is:\\ 
    \verb+1 RNN layer, 3 transformer layers; 1 RNN layer +  (Willow experiments) or \\
    \verb+1 RNN layer, 2 transformer layers; 1 RNN layer +  (Scaling experiments).
    \item The readout layer has 2 spatial mixing transformer layers before the readout network described above. 
    \item The internal representation per stabilizer has 128 channels  (256 for Willow experiments).
    \item The RNN layers use a gated recurrence\cite{Griffin}, which is faster because it only involves element-wise operations (Fig.~\ref{fig:griffin}).
    \item  For the real-time colour code model we added gating to the attention~\cite{qiu2025gated}.
\end{itemize}
For the Willow data which has durations with multiples of 10 cycles, we temporally compressed groups of 5 cycles instead of 6.

\subsection{Training}
While AQ1 used a combination of pretraining on synthetic noise and fine-tuning on experimental data (simulated for larger code distances), since the focus here is on scaling to surface code sizes well beyond current hardware, we conduct the scaling experiments using SI1000 and train in a single phase without the need for fine-tuning. For the Willow data, we follow the same protocol as used in the Willow paper of training five models for each code distance on SI1000, fine-tuning on samples taken from a Detector Error Model (DEM) fitted to experimental data, followed by a final fine-tuning using the limited experimental samples while holding out a test set of experimental samples (as well as an evaluation set to monitor fine-tuning progress). 

We train our models on examples from a range of code-distances. A fully-trained model is able to decode examples from any of the distances seen in training. For ease of training, we follow a code distance curriculum, where we change the distribution of samples over time, initially focusing on the smaller code distances and finishing by focusing on larger code distances. While this allows us to more quickly achieve high accuracy for larger code distances, and avoids separate training for each code distance, maintaining performance across all code distances does involve sacrificing a small amount of accuracy. We train two models for each code --- one only up to distance 11 and the other for higher distances. To achieve the best performance at distances  23 and 27, we conduct final training only on examples for a single distance, resulting in a network specialized for just that code distance. To extract the absolute best performance, it would be possible to carry out such fine-tuning for each code distance, but here we just focus on the highest code distances where the focused training makes the greatest difference.

We note that for the surface code at distance 23, we achieve an LER of around $7.3 \times 10^{-11}$, but only see $3\times10^9$ training examples or $3.6 \times 10^{11}$ cycles, so we expect to see very few incorrect examples in the latter half of training, but the cross-entropy objective leads to beneficial gradients from the correctly classified examples. 

\subsubsection{Noise}
While for all the simulations we target a noise level of p=0.15\%, we find it beneficial to train with a mixture of harder and easier examples, and it is easier for the model to learn with examples with a lower noise level. We generate each training example with a noise level sampled from a distribution (see Table~\ref{tab:hyperparameters}). 

\subsubsection{Masking}
We find that training the model with dropout on the inputs (akin to that used in BERT\cite{devlin2019bert}) leads to better results. We drop out 50\% of stabilizer representations in each cycle, replacing them with zero-vectors. We do so on only 80\% of training examples to ensure that the model sees examples with all stabilizer representations (as are used in inference).

\subsubsection{Auxiliary losses}
During training we predict a number of auxiliary losses to improve the speed of conversion and to push the network to decode at any shot length (See \ref{sec:intermediate_data}). As in AQ1 we can predict multiple logical observables for the logical qubit for any experiment, and by simulating terminating the experiment at any cycle, with a set of data qubit measurements at each cycle, we train the network to predict these logical observables at every cycle.  Furthermore, we simulate such logical observables in each basis but using separate parameters in the readout network.  

\subsubsection{Code distance curriculum}
\begin{figure}[t]
\centering
\includegraphics[width=\textwidth]{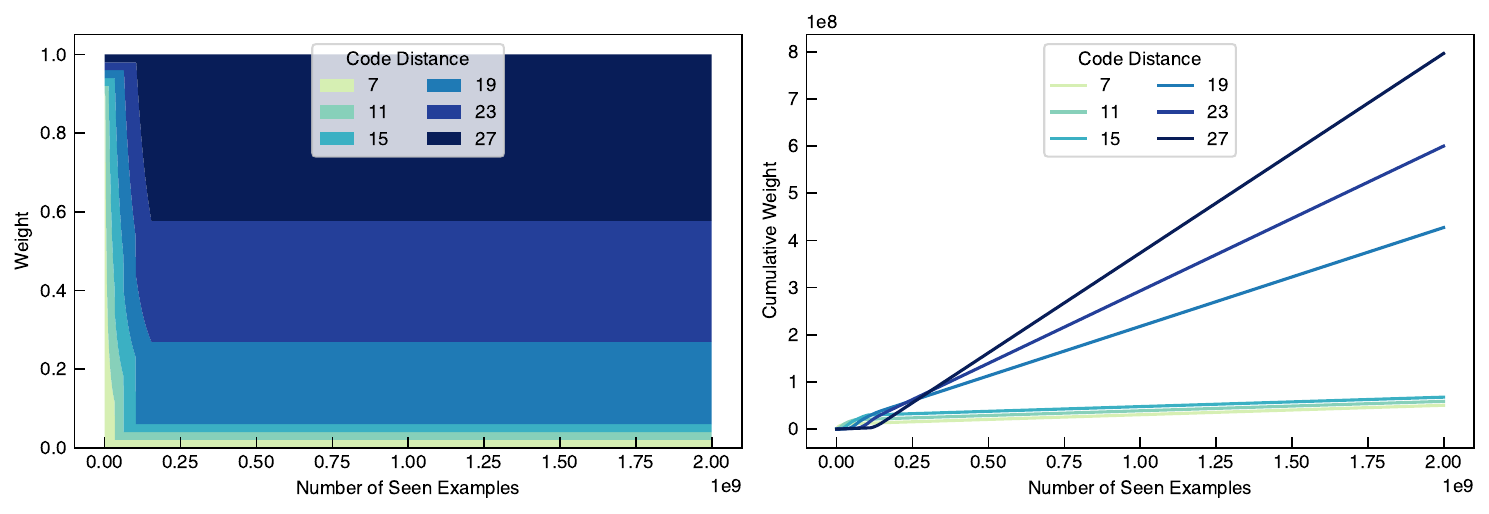}
\caption{
\textbf{Code distance curriculum.} Shown are weights (left) and the cumulative number of samples seen (right), over a period of $2\times 10^9$ total training samples.}
\label{fig:cc_curriculum}
\end{figure}

We use a code distance curriculum which gradually increases the weight on larger code distances as training progresses. The weights and cumulative number of samples seen over a 2B training run can be seen in Fig.~\ref{fig:cc_curriculum}.

\subsubsection{Optimization}
We generally use the Lion optimizer\cite{lion}. For fine-tuning we adopted the Muon optimizer \cite{muon}. We use a batch size of 1024 throughout, apart from the finetuning runs on Willow data for which we used a batch size of 64 ($d=3$), 128 ($d=5$), or 256 ($d=7$).

\subsubsection{Learning rate}
We have a 1-million example linear warm-up period to reach the full learning rate and use a fixed learning rate throughout training with an optional cosine decay, particularly for fine-tuning (\cref{tab:hyperparameters}, \cref{tab:ensemble_hyperparameters}).

We scale the learning rate for each batch according to the code distance and number of cycles with the following formula:

\[\mathrm{learning\_rate} = \mathrm{base\_learning\_rate} \times 0.8^{\log_2 N_s/8} \times 2 ^ {\log_2 R/24}\]
where $N_s$ is the number of stabilizers in the example at the given code distance and $R$ is the number of cycles. (Double the number of cycles for the Bell-flagged colour code).

\begin{table}
\centering
\footnotesize
\begin{tabular}{p{40mm}|p{24mm}|p{24mm}|p{24mm}|p{24mm}}
\toprule
Parameter & \multicolumn{2}{c|}{Surface} & \multicolumn{2}{c}{Bell-flagged Colour} \\
\midrule
Training code distances & 3, 5, 7, 9, 11 & 7, 9, 11, 13, 15, 17, 19, 21, 23 & 3, 5, 7, 9, 11 & 7, 11, 15, 19, 23, 27\\
\midrule
Evaluation code distances & 3, 5, 7, 9 & 11, 15, 19, 23 & 3, 5, 7, 9, 11 & 15, 19, 23, 27\\
\midrule
cycles for training & 24, 48, 72, 120 & 24, 48, 72, 120 & 120 & 120 \\
\midrule
Parameters & 32,177,930 & 32,177,930 & 29,557,106 & 29,539,186 \\
\midrule
Channels & 512 & 512 & 512 & 512 \\
\midrule
Heads / key size / value size & 16 / 32/ 32 & 16 / 32 / 32 & 16 / 32 / 32 & 16 / 32 / 32 \\
\midrule
Widening & 4 & 4 & 4 & 4 \\
\midrule
Pooling transformer layers / heads / key \& value size & 4 / 8 / 32 & 4 / 8 / 32 & 2 / 8 / 32 & 2/8/32 \\
\midrule
Total training examples & 2e9 & 2e9 & 2e9 & 7.0e8 \\
\midrule
\makecell[tl]{Noise distribution\\($p$, relative weight)} & \makecell[tl]{(0.001, 1)\\(0.0015, 5)\\ (0.002, 2)\\(0.0025, 3)\\(0.003, 3)\\(0.0035, 2)\\(0.004, 2)} & 
\makecell[tl]{(0.001, 1)\\(0.0015, 5)\\(0.002, 2)\\(0.0025, 3)\\(0.003, 3)\\(0.0035, 2)\\(0.004, 2)} & 
\makecell[tl]{(0.001, 1)\\(0.0015, 1)\\(0.002, 4)\\(0.0025, 3)\\(0.003, 2)\\(0.0035, 1)} & \makecell[tl]{(0.001, 1)\\(0.0015, 6)\\(0.002, 4)\\(0.0025, 3)\\(0.003, 2)\\(0.0035, 1)}\\
\midrule
Learning rate & 2e-5 & 2.2e-5 & 2.4e-5 & 2.5e-5 \\
\bottomrule
\end{tabular}
\caption{\textbf{Model and training hyperparameters for the four base models.}}
\label{tab:hyperparameters}
\end{table}

\begin{table}
\footnotesize
\begin{tabular}{p{40mm}|p{24mm}|p{24mm}p{24mm}p{24mm}}
\toprule
Parameter & Surface & \multicolumn{3}{c}{Bell-flagged Colour} \\
\midrule
Code distances & 23 & 19 & 23 & 27\\
\midrule
Cycles for training & 72, 120, 168 & 120 & 72, 120, 168 & 120 \\
\midrule
Total training examples & $3\times10^9$ & $9.5\times10^8$ & $3\times10^9$ & $1.3\times10^9$ \\
\midrule
Muon learning rate & $2.5\times10^{-3}$ & \makecell[tl]{$4\times10^{-4}$\\$1\times10^{-3}$} & $2.5\times10^{-3}$ & \makecell[tl]{$1.0\times10^{-3}$\\ $1.5\times~10^{-3}$\\ $2\times10^{-3}$} \\
\midrule
Ensemble & Base model with 1 finetune & 2 finetunes & 2 finetunes & 3 finetunes\\
\midrule
Ensemble $\log_{10}$ LER & $-10.135$ & $-8.052$ & $-9.456$ & $-10.398$\\
\midrule
Model $\log_{10}$ LERs &  $-10.079$  $-9.778$& \makecell[tl]{$-7.714$\\$-7.764$} & $-8.589$  $-8.530$  & \makecell[tl]{$-8.127$\\$-8.233$\\$-8.474$} \\
\midrule
Learning rate decay factor & $0.2$ & $0.25$  & $0.1$ & $0.25$  \\
Decay examples range &  2e9 to 3e9 &  5.6e8 to 1e9 & 7e8 to 2.3e9 & 5.6e8 to 1e9 \\
\bottomrule\end{tabular}
\caption{\textbf{Fine-tuning hyperparameters for the variants used in ensembles.}}
\label{tab:ensemble_hyperparameters}
\end{table}

\subsection{Data}

\begin{figure}[t]
\centering
\includegraphics[width=0.8\textwidth]{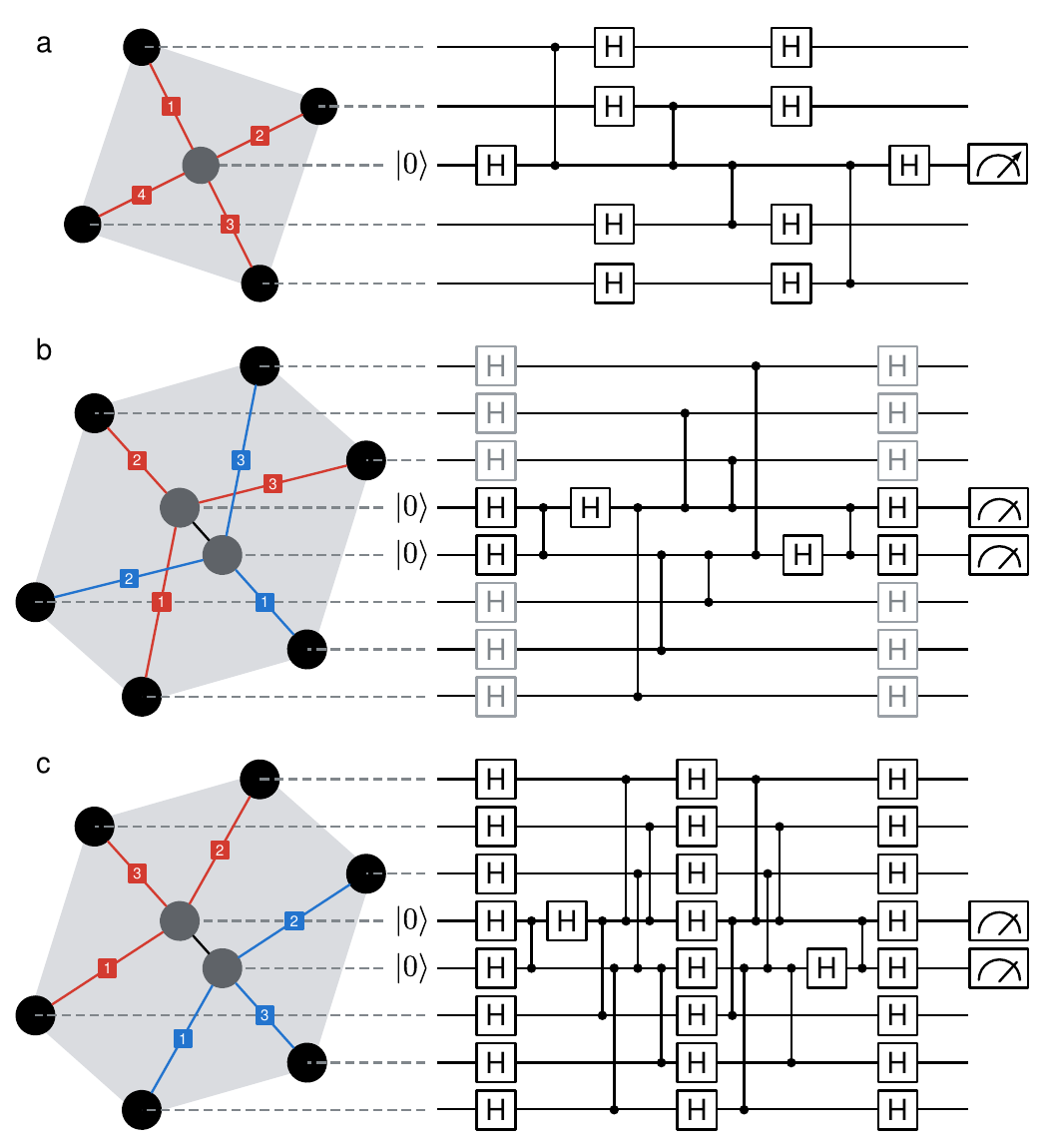}
\caption{
\textbf{Readout circuits and topology for surface and colour codes}. \textbf{a,} Common X and Z stabilizer readout for the XZZX rotated surface code\cite{Bonilla_Ataides2021}. For an adjacent plaquette, the readout order is anti-clockwise s.t.\ data qubits are touched in a consistent order for overlapping stabilizers. \textbf{b,} Stabilizer readout for Bell-flagged colour code\cite{Baireuther2019}, where the grey Hadamard gates are applied for an $X$ readout, and not for a $Z$ readout. The topology is non-planar as the $CZ$ gates applied between ancilla and data qubits cross. \textbf{c,} Joint $X$ and $Z$ stabilizer readout for superdense colour code\cite{gidney2023chromobius}. This topology is planar.}
\label{fig:readout-circuits}
\end{figure}
For our standard evaluations we chose 120-cycle memory experiments to have durations longer than $d \times 3$ for all code distances, and retaining flexibility in choosing the group length for temporal compression.

\subsubsection{Circuits}
All surface code data is generated from a XZZX surface code with the same readout ordering as in our previous work \cite{Bausch2024}, see \cref{fig:readout-circuits}a.

As we are targeting a superconducting hardware with AQ2-RT, and to enable running the Chromobious decoder as a comparison, our superdense circuits follow a planar readout order inspired by Gidney et al.\cite{gidney2023chromobius} (\cref{fig:readout-circuits}c). 

We generally use the Bell-flagged colour code for most of our analysis. However, due to the impossibility of running Chromobius on our Bell-flagged code implementation, for the real-time analysis we use superdense colour code circuits with planar readout of the same size and noise model. We directly compare the full AQ2 model performance on both superdense and Bell-flagged variants in \cref{fig:realtime}.

\subsubsection{Circuit level noise}
We use SI1000\cite{Gidney2022} (superconducting-inspired 1,000-ns cycle duration) noise in our simulations. (See Figs.~\ref{fig:surface-code-circuits}, \ref{fig:color-code-circuits} and \ref{fig:superdense-circuits} for full circuit details.) 
SI1000 is a circuit depolarising noise model that approximates the noise processes occurring in superconducting circuits. We note that there was some ambiguity in the original definition\cite{Gidney2021honeycomb} of SI1000, now resolved\cite{gidney2023yoked}. In this work we fully align with the modern interpretation, which is different from the SI1000 interpretation used in our previous work\cite{Bausch2024} in two ways:
Both \(\mathrm{Idle}(p)=p/10\) and \(\mathrm{ResonatorIdle}(p) = 2p\) noise terms are added after measurements and resets (instead of just \(\mathrm{ResonatorIdle}(p)\)).
Measurement and reset count as two separate operations with full sets of noise applied (instead of a joint operation with a single set of noise)
This produces a minor increase in events at each particular value of $p$.

\newcommand{\circuitsvgs}[5]{%
    $\underbrace{%
        \foreach \i in {#2,...,#3}{%
          \includegraphics[scale=#5]{figures/circuits/circuit-#1--i=\i.pdf}%
          \hspace{2pt}%
        }%
    }_{\text{#4}}$\hspace{-2pt}%
}

\begin{figure}[t]
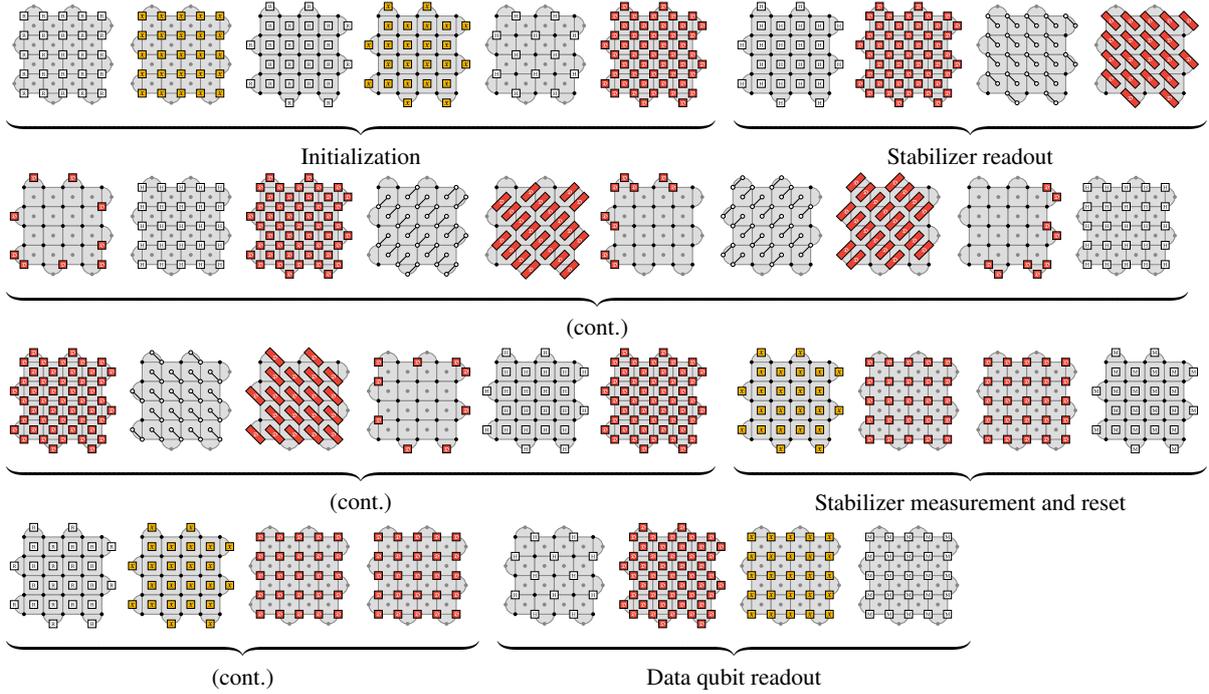

    \circuitsvgs{sc}{0}{5}{Initialization}{0.17}
    \circuitsvgs{sc}{6}{9}{Stabilizer readout}{0.17}\\
    \circuitsvgs{sc}{10}{19}{(cont.)}{0.17}\\
    \circuitsvgs{sc}{20}{25}{(cont.)}{0.17}
    \circuitsvgs{sc}{26}{29}{Stabilizer measurement and reset}{0.17}\\
    \circuitsvgs{sc}{30}{33}{(cont.)}{0.17}
    \circuitsvgs{sc}{34}{37}{Data qubit readout}{0.17}
    \caption{\textbf{Circuit sequence for the surface code.} Bit flip errors are denoted in yellow, depolarizing channels in red. CZ gates are marked with their circuit symbol. Measurement, reset, and Hadamard gates are marked with an M, R, and H, respectively.}
    \label{fig:surface-code-circuits}
\end{figure}

\begin{figure}[t]
    \circuitsvgs{cc}{0}{1}{Initialization}{0.125}
    \circuitsvgs{cc}{2}{8}{Stabilizer readout (X)}{0.125}\\
    \circuitsvgs{cc}{9}{17}{(cont.)}{0.125}\\
    \circuitsvgs{cc}{18}{24}{(cont.)}{0.125}
    \circuitsvgs{cc}{25}{26}{Stabilizer measurement}{0.125}\\
    \circuitsvgs{cc}{27}{32}{(cont.)}{0.125}
    \circuitsvgs{cc}{33}{35}{Stabilizer readout (Z)}{0.125}\\
    \circuitsvgs{cc}{36}{44}{(cont.)}{0.125}\\
    \circuitsvgs{cc}{45}{53}{(cont.)}{0.125}\\
    \circuitsvgs{cc}{54}{55}{(cont.)}{0.125}
    \circuitsvgs{cc}{56}{62}{Stabilizer measurement}{0.125}\\
    \circuitsvgs{cc}{63}{63}{(cont.)}{0.125}
    \circuitsvgs{cc}{64}{65}{Data qubit readout}{0.125}
    \caption{\textbf{Circuit sequence for the Bell-flagged colour code.} Bit flip errors are denoted in yellow, depolarizing channels in red. CZ gates are marked with their circuit symbol. Measurement, reset, and Hadamard gates are marked with an M, R, and H, respectively.}
    \label{fig:color-code-circuits}
\end{figure}

\begin{figure}[t]
    \circuitsvgs{ccsd}{0}{1}{Initialization}{0.125}
    \circuitsvgs{ccsd}{2}{8}{Stabilizer readout}{0.125}
    \circuitsvgs{ccsd}{9}{17}{(cont.)}{0.125}
    \circuitsvgs{ccsd}{18}{26}{(cont.)}{0.125}
    \circuitsvgs{ccsd}{27}{35}{(cont.)}{0.125}
    \circuitsvgs{ccsd}{36}{44}{Stabilizer measurement}{0.125}
    \circuitsvgs{ccsd}{45}{46}{Data qubit readout}{0.125}
    \caption{\textbf{Circuit sequence for the superdense colour code.} Bit flip errors are denoted in yellow, depolarizing channels in red. CZ gates are marked with their circuit symbol. Measurement, reset, and Hadamard gates are marked with an M, R, and H, respectively. The controlled-X gates during stabilizer measurement are displayed for completeness and not actually executed on the device. Instead, one can classically track this feedback by updating the Pauli frame.}
    \label{fig:superdense-circuits}
\end{figure}

Example noise in measurement-and-reset stabilizer qubits in old and new SI1000 interpretation. Before (AQ1\cite{Bausch2024}):

\begin{verbatim}
X_ERROR(0.01) 7 8 9 10 11 12      # 5p pre-measurement noise
MR 7 8 9 10 11 12
X_ERROR(0.004) 7 8 9 10 11 12     # 2p post-measurement noise
DEPOLARIZE1(0.004) 0 1 2 3 4 5 6  # 2p ResonatorIdle noise
\end{verbatim}

In this work:

\begin{verbatim}
X_ERROR(0.01) 7 8 9 10 11 12       # 5p pre-measurement noise
M 7 8 9 10 11 12
DEPOLARIZE1(0.0002) 0 1 2 3 4 5 6  # p/10 further measurement idle noise
DEPOLARIZE1(0.004) 0 1 2 3 4 5 6   # 2p Idle noise (due to measurement)
R 7 8 9 10 11 12
X_ERROR(0.004) 7 8 9 10 11 12      # 2p Post-measurement noise
DEPOLARIZE1(0.0002) 0 1 2 3 4 5 6  # p/10 Idle noise
DEPOLARIZE1(0.004) 0 1 2 3 4 5 6   # 2p ResonatorIdle noise (due to reset)
\end{verbatim}

\subsubsection{Intermediate data}
\label{sec:intermediate_data}
When simulating data, we have some privileged information about quantum states, which would be inaccessible in a real experiment due to the no-cloning theorem. We use this privileged information to provide the model with some auxiliary labels at each experiment step. There are two types of intermediate auxiliary information.

\paragraph{[fake endings].} Fake alternative endings of the experiment at every cycle. We calculate them by measuring the data qubits and deriving a fake alternative last cycle of stabilizer measurements and observables. 

\begin{figure}[t]
\centering
\includegraphics[width=0.7\textwidth]{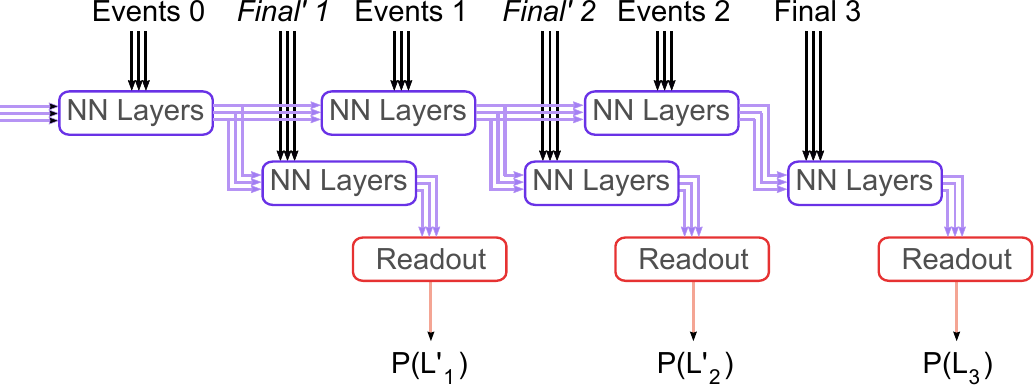}
\caption{
\textbf{Fake endings.} By simulating early terminations of an experiment we can measure final data qubits, give the network the corresponding final stabilizers and thus predict a logical observable $L'$ for each cycle of an experiment in addition to the final logical observable $L$.}
\label{fig:fake_intermediates}
\end{figure}

\paragraph{[noiseless observables].} We use MPP (multi-pauli instructions) measurements on the data qubits in the observable line, to obtain a noiseless measurement of the observable at each cycle. We entangle it in a Bell pair to provide a result also for an orthogonal measurement basis.

We calculate each intermediate observable for a number of equivalent measurement lines. For the surface code we choose each vertical (resp. horizontal for X experiments) line of data qubits, giving us code-distance bits per intermediate observable. For the colour code we chose the data qubits in each of the three sides of the triangular patch, giving us 3 bits per intermediate observable.

We combine these two sources of intermediate measurements to provide the model with 4 intermediate labels at each cycle:
\paragraph{[fake intermediates].} We use an extra head at each group of cycles that receives the fake ending stabilizer measurements to predict the fake ending observables.
\paragraph{[noiseless].} An extra head at each time step predicting the noiseless observable at each time chunk
\paragraph{[noiseless delta].} An extra head at each time step predicting the xor (exclusive or) between two noiseless observables at the beginning and end of each time chunk.
\paragraph{[noiseless to intermediate delta].} An extra head at each time step that receives the fake ending stabilizer measurements to predict the xor between a noiseless observable at the beginning of the time chunk and a fake ending observable at the end of the time chunk.

Each label is used in optimization as an auxiliary binary cross-entropy loss. During training we balance these auxiliary losses and the final loss by optimising a weighted sum of the losses (see weights in Table~\ref{tab:lossweights}). During evaluation, we do not need to make predictions from any auxiliary heads. To minimize bugs, the held-out test data used to calculate the final reported accuracies is in all cases calculated from a simple circuit and data pipeline that never generates any intermediate measurements nor labels.

\subsubsection{Unified data representation}
\label{sec:unified}
In the Bell-flagged colour code, stabilizer and flag qubits  are measured in alternating bases in two sub-cycles. Thus the measurements at a given physical location have two different meanings. To make a representation more consistent with that of the surface code, with spatial consistency, we reshape the two sub-cycles into a single cycle representation. The X and Z basis stabilizer measurements are interleaved, to make a unified representation for the cycle. The flag measurements are similarly interleaved, aligned with the stabilizer measurements. When using the unified representation we combine only 3 consecutive cycles of information before presentation to the neural network.

\begin{table}
\centering
\begin{tabular}{l|l}
\toprule Loss & Weight \\ 
\midrule 
Final & 1.2 \\ 
Fake intermediate & 1 \\
Noiseless & 1 \\
Noiseless delta & 1 \\
Noiseless to intermediate delta & 8 \\
\bottomrule 
\end{tabular}
\caption{Loss weights}
\label{tab:lossweights}
\end{table}

\subsubsection{Willow experimental data}
For the experimental data experiments we use the data from the Willow paper \cite{Willow} and follow the same protocol. The data consists of 50,000 shots of training data for each duration \{10, 30, 50, … 250\} cycles for each distance \{3, 5, 7\} and in each basis \{X, Z\}. Distance 3 and 5 datasets are taken at each of 9 and 4 locations, respectively, on the 105-qubit Willow chip. A Detector Error Model (DEM) fitted to data from a separate 13-cycle dataset at each distance was used for the initial fine-tuning, and an RL-modified DEM was used by the matching decoder. For the final fine-tuning we do two-fold cross validation, splitting each set into two parts. One half is used for final evaluation and the other split into 19,880 fine-tuning examples and 5,120 evaluation examples to measure the progress of fine-tuning. 

\subsection{Measuring accuracy}
Accuracy is measured with logical error per cycle (LER). 

For evaluating on shots of a single length $n$ cycles, of which a fraction $E(n)$ are incorrectly decoded, we compute $\mathrm{LER} = {1 \over 2} (1 - \sqrt[n]{1-2E(n)})$. For the experimental data, with test data of various durations, we fit a straight line to the log fidelities \[F(n) := 1 - 2E(n):\]
\[\log F(n) = \log F_0 + n\log(1-2\epsilon),\] following \cite{Willow}.

A cycle is defined here to involve the measurement of the stabilizers in both bases, so for the Bell-flagged colour code this involves measuring all the stabilizer and flag qubits twice (for X and Z basis sequentially, \cref{fig:color-code-circuits}). Therefore with the same clock-speed a cycle in the colour code will take about twice as long as for the surface code or the superdense colour code. This is an approximation, as the readout circuits are different, for instance associating each stabilizer with 3 data qubits in the colour code and 4 in the surface code (\cref{fig:readout-circuits}).

We measure the accuracy of the decoders on a fixed, held-out dataset of examples from Stim. These datasets were only used for final evaluation of the selected model of each configuration. Since higher code-distance decoders have lower error rates, we require larger test sets to accurately estimate the error rates. The number of samples used for each code distance for each code is shown in \cref{tab:num_samples}. 
We used independent large benchmarking datasets of the same sizes to compare between models trained with different hyperparameters and to make the final selection of which model to evaluate with the held-out test set. 

\begin{table}
    \centering
\begin{tabular}{c|rr}
\toprule
\textbf{Code distance} & \textbf{Samples for Surface Code} & \textbf{Samples for Color code} \\
\midrule
3 & 500,000 & 500,000 \\
\midrule
5 & 2,000,000 & 1,500,000 \\
\midrule
7 & 5,000,000 & 3,000,000 \\
\midrule
11 & 10,000,000 & 6,000,000 \\
\midrule
15 & 50,000,000 & 25,000,000 \\
\midrule
19 & 250,000,000 & 250,000,000 \\
\midrule
23 & 2,500,000,000 & 500,000,000 \\
\bottomrule
\end{tabular}
  \caption{\textbf{Numbers of samples used in evaluations.} For the SI1000 0.15\% benchmarking and final, held-out test sets, by code distance (for 120-cycle shots).}
    \label{tab:num_samples}
\end{table}

The number of samples was chosen to achieve reasonable error bars for the LER at the expected LER for each code distance.
During development we use separate temporary development datasets to monitor the training of the networks and determine when to finish training. Evaluation on datasets of the above sizes is too expensive, particularly for larger code distances where examples are more expensive to evaluate, as well as requiring larger test sets to estimate error rates accurately. To achieve faster evaluation with sufficiently low standard error, we evaluate on data with much higher noise, for which the error rate is correspondingly higher and evaluation requires fewer samples. For the surface and colour codes we use $p=0.3\%$ and $p=0.25\%$ noise respectively. While improving performance on the high noise data is no guarantee of improving performance at the target 0.15\% noise level, this is a useful proxy which makes preliminary model comparison much less compute-intensive.

\subsubsection{Long experiment datasets}
To benchmark surface code decoders on longer experiments, we decrease the number of samples used in proportion to the increase in duration (Table~\ref{tab:num_samples_long}).  Numbers of rounds are chosen to be multiples of 24 (using a group size of 6 cycles and a block size of 24 or 48). 

\begin{table}
    \centering
    \begin{tabular}{c|rrrrr}
    \toprule
Code distance  & 
\multicolumn{5}{c}{cycles}\\
    \cline{2-6}
&
120 & 
1,080 & 
10,800 & 
108,000 & 
1,080,000\\
\midrule
11 & 
10,000,000 & 
1,125,000 & 
125,000 & 
25,000 & 
25,000\\
15 & 
50,000,000 & 
5,575,000 & 
575,000 & 
75,000 & 
25,000\\
19 & 
250,000,000 & 
27,800,000 & 
2,800,000 & 
300,000 & 
50,000 \\
23 & 
2,500,000,000 & 
277,800,000 & 
27,800,000 & 
2,800,000 & 
300,000     \\
\bottomrule
\end{tabular}
    \caption{\textbf{Numbers of samples used in evaluations.} For the SI1000 0.15\% benchmarking and final, held-out test sets for long experiment evaluations, according to number of cycles.}
    \label{tab:num_samples_long}
\end{table}

\subsubsection{Low noise samples 0.1\% noise}
To measure error rates at 0.1\% noise, we use $10\times$ more samples than for 0.15\% noise (Table~\ref{tab:lownoisesamples}).
\begin{table}
    \centering
    \begin{tabular}{r|rr}
\toprule
Code distance &
Samples for Surface Code &
Samples for color code \\
\midrule
3 &
5,000,000 &
5,000,000 \\
5 &
20,000,000 &
15,000,000 \\
7 &
50,000,000 &
30,000,000 \\
11 &
100,000,000 &
60,000,000 \\
15 &
500,000,000 &
250,000,000 \\
19 &
 &
2,500,000,000 \\
\bottomrule
\end{tabular}
    \caption{\textbf{Numbers of samples used in held-out test set for 0.1\% noise.}}
    \label{tab:lownoisesamples}
\end{table}

\subsubsection{Comparing AQ2 across codes}

\begin{figure}[t]
\centering
\includegraphics[width=0.7\textwidth]{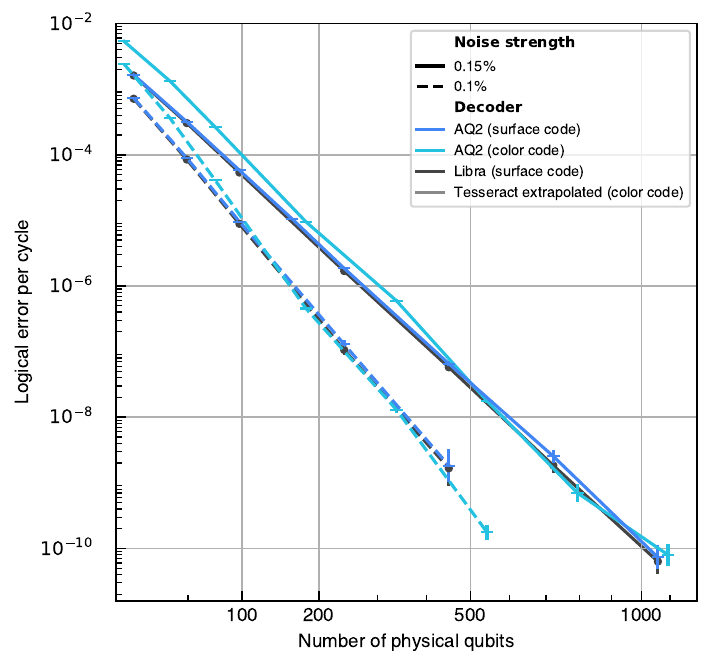}
\caption{
\textbf{Comparison between codes.} Logical error per cycle versus number of physical qubits for AQ2, comparing surface and Bell-flagged colour code decoding at $0.15\%$ and $0.1\%$ noise, together with Libra and (at $0.1\%$) extrapolated Tesseract baselines.}
\label{fig:accuracy_point1}
\end{figure}

By plotting the logical error rates of \cref{fig:accuracy}a \& b, as well as the corresponding rates for $0.1\%$ noise, against the number of physical qubits required to implement the different codes, we can see (\cref{fig:accuracy_point1}) that the accuracies for colour and surface code at these noise levels are comparable. The decoders scale well at the lower noise level, matching the accuracy of Libra on the surface code and coming close to the extrapolated Tesseract error rate, confirming the noise generalization shown in \cref{fig:longrounds}b. 

\subsection{Measuring speed}
\label{sec:measuring-speed}
We measure throughput and latency of AQ2 models on a Trillium TPU. The model is executed in a streaming fashion where the decoder accumulates a block of cycles before processing them jointly. While the model is executed on a block, new cycles are accumulated into a buffer. At the end of the memory experiment, the readout network is executed to get the final prediction. In order to measure throughput, we measure the duration from the experiment start to when the final prediction is complete and divide by the number of cycles:
 \[\text{throughput} := \frac{\text{duration}}{\text{num\_cycles}} \]
We choose a large number of cycles (100,800) to approximately cancel out constant effects on timing such as executing the final readout network. All timing is carried out with a single memory experiment at a time, but we note that, for lower code distances, multiple logical qubits can be decoded in parallel on a single device within the throughput, amortizing the compute hardware requirement.

We measure latency by introducing a simulated clock rate such that the decoder receives the measurements of a single cycle at a fixed frequency, e.g. every $1~\upmu s$. The latency of a decoder is then defined as the delay between receiving the last cycle and the prediction being ready:
 \[\text{latency} := \text{time\_prediction\_done} - \text{time\_last\_cycle} \]

\begin{figure}[t]
    \centering
    \captionsetup[subfigure]{position=top, skip=0pt}
    \begin{subfigure}[b]{0.9\linewidth}
        \centering
        \caption{}
        \includegraphics[width=\linewidth]{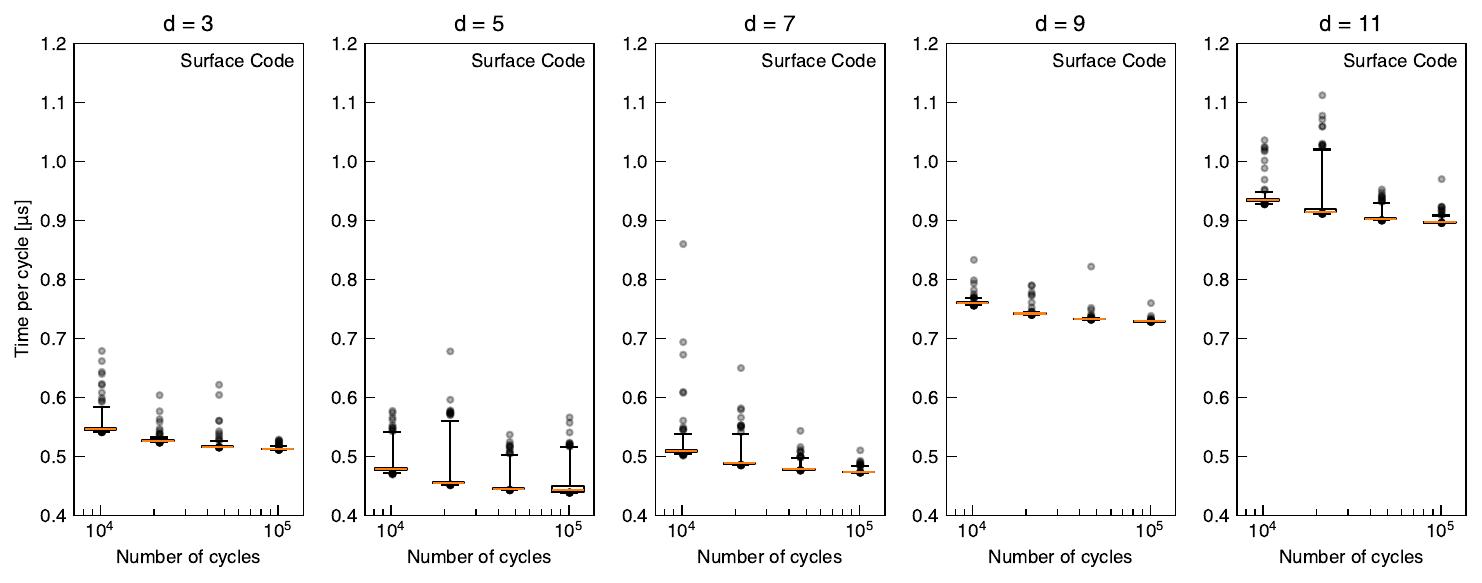}
        \label{fig:throughput}
    \end{subfigure}
    \hfill%
    \begin{subfigure}[b]{0.9\linewidth}
        \centering
        \caption{}
        \includegraphics[width=\linewidth]{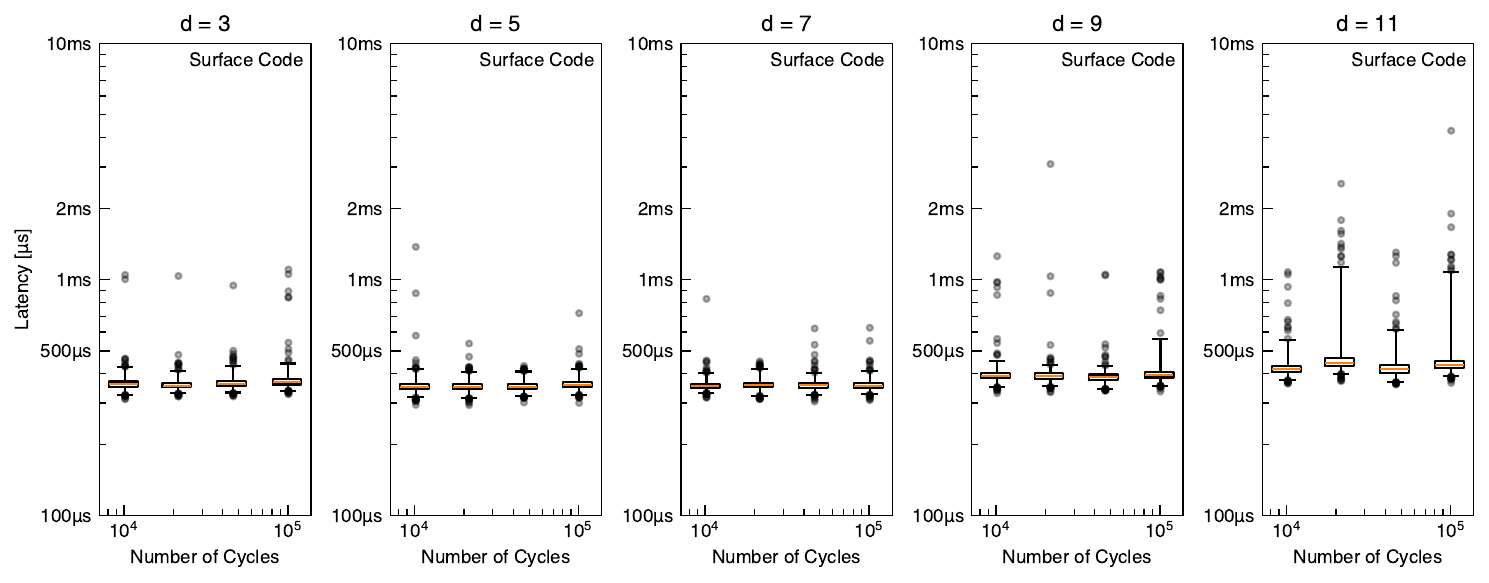}
        \label{fig:latency}
    \end{subfigure}
    \hfill
    \caption{\textbf{Throughput and latency measurements for surface code.}
    Box plots of decoding time per cycle \textbf{(a)} and latency \textbf{(b)} for different code distances and memory experiment lengths. Each plot represents 1000 shots of 100,800 cycles with whiskers extending from 1st to 99th percentile. Latencies are measured with a simulated cycle time of $1\upmu s$.}
    \label{fig:throughputs}
\end{figure}

\begin{figure}[t]
    \centering
    \captionsetup[subfigure]{position=top, skip=0pt}
    \begin{subfigure}[b]{0.8\linewidth}
        \centering
        \caption{}
        \includegraphics[width=\linewidth]{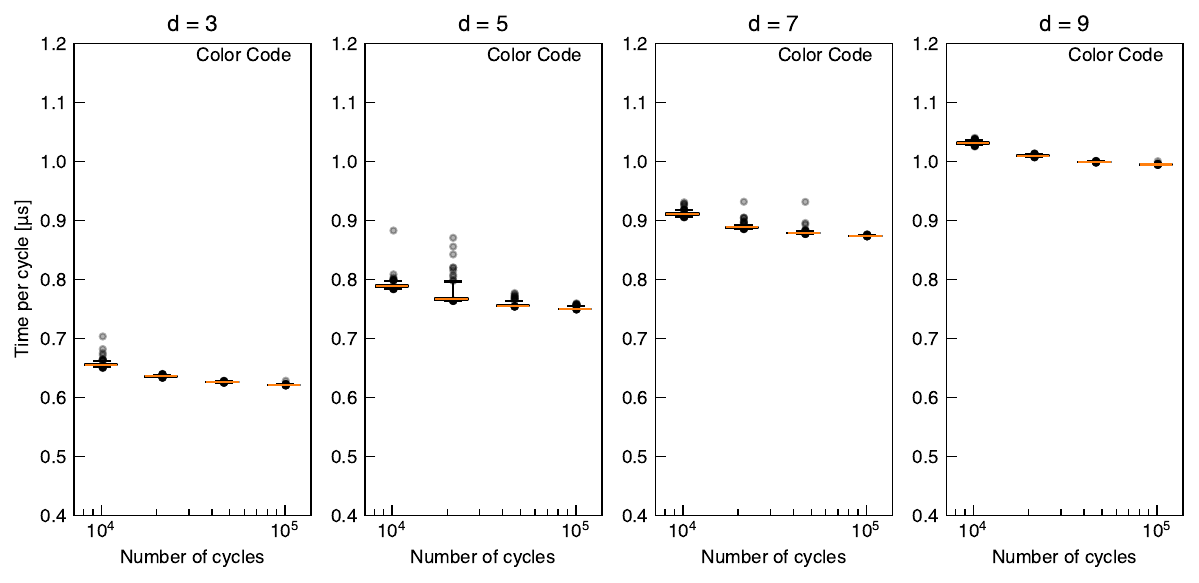}
        \label{fig:throughput-cc}
    \end{subfigure}
    \begin{subfigure}[b]{0.8\linewidth}
        \centering
        \caption{}
        \includegraphics[width=\linewidth]{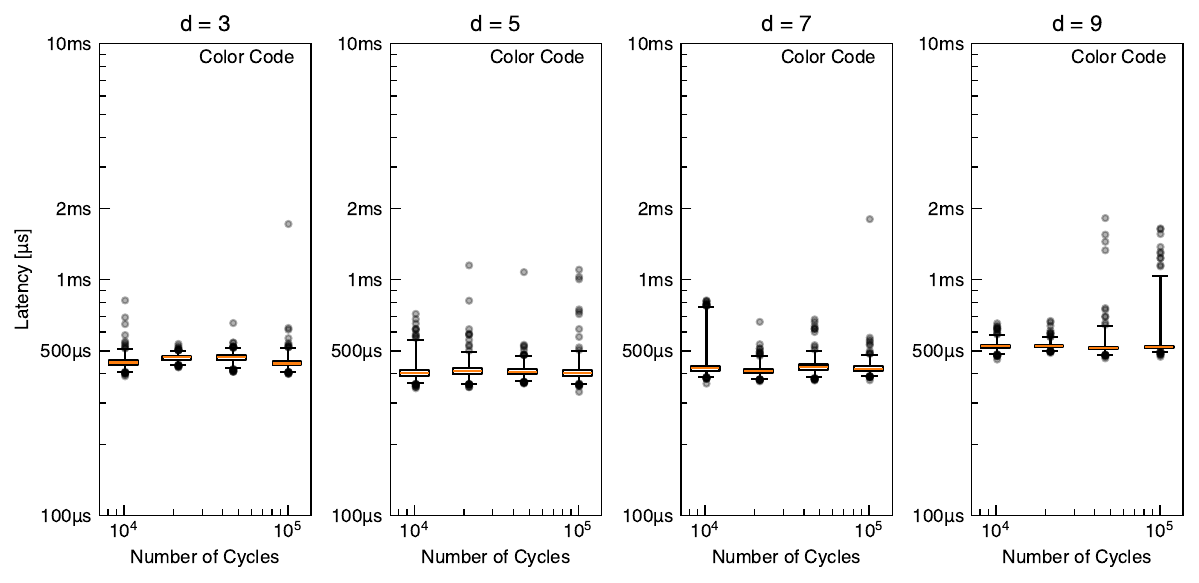}
        \label{fig:latency-cc}
    \end{subfigure}
    \caption{\textbf{Throughput and latency measurements for superdense colour code.}
    Box plots of decoding time per cycle \textbf{(a)} and latency \textbf{(b)} for different code distances and memory experiment lengths. Each plot represents 1000 shots of 100,800 cycles with whiskers extending from 1st to 99th percentile. Latencies are measured with a simulated cycle time of $1\upmu s$.}
    \label{fig:throughputs-cc}
\end{figure}

The latency of AQ2-RT remains constant for increasing numbers of cycles\cite{Willow} (\cref{fig:throughputs}b and \cref{fig:throughputs-cc}), showing that the decoder can keep up with the simulated clock rate of $1~\upmu s$. It should be noted that the throughput and latency of AQ2 are independent of the detection fraction and the error pattern, unlike matching-based decoders whose runtime increases with error density and can have extreme worst-case decoding times that necessitate lower-accuracy fall-back mechanisms.

\begin{figure}[t]
    \centering
    \captionsetup[subfigure]{position=top, skip=0pt}
    \begin{subfigure}[b]{0.8\linewidth}
        \centering
        \caption{}
        \includegraphics[width=0.9\linewidth]{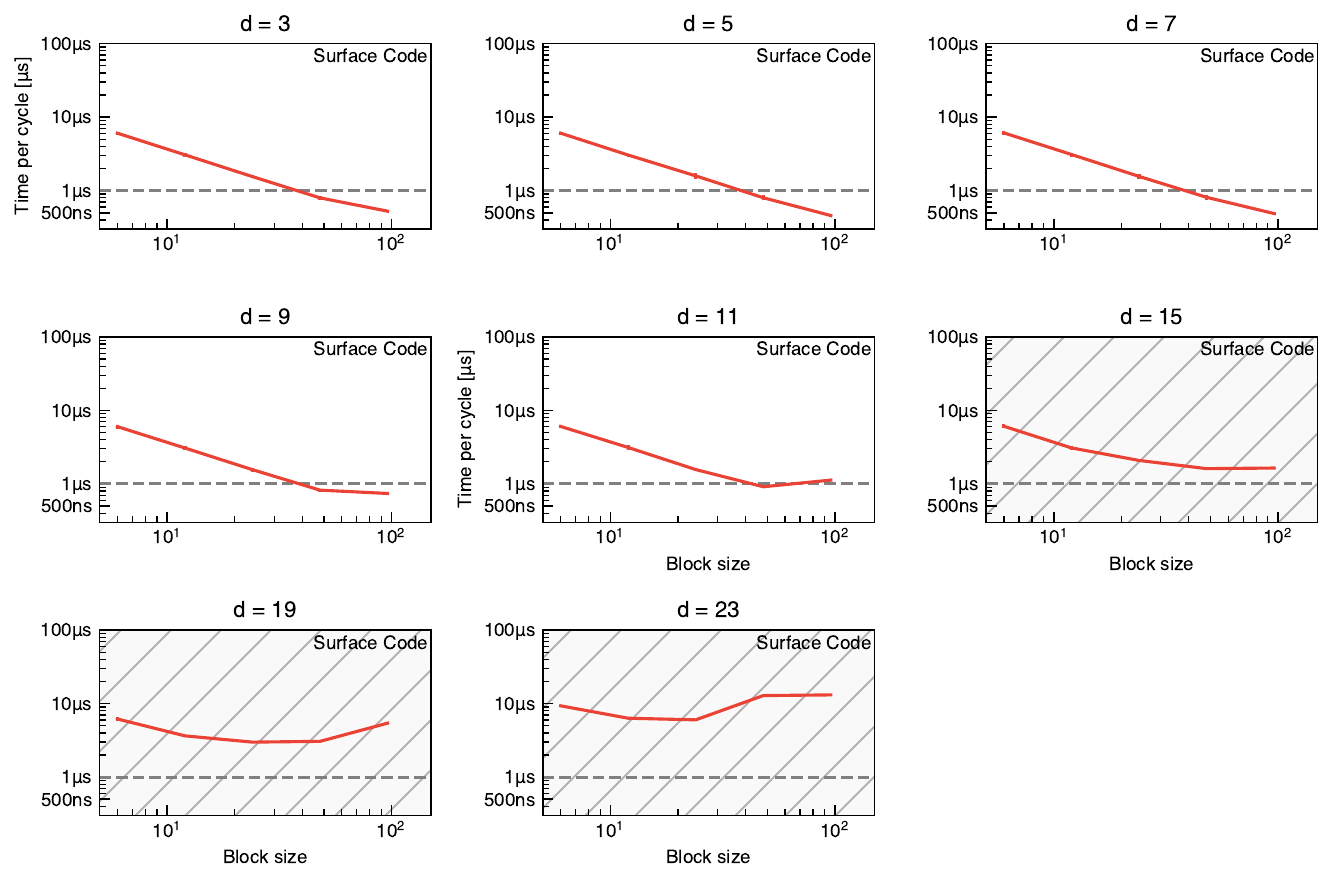}
        \label{fig:blocksize}
    \end{subfigure}
    \begin{subfigure}[b]{0.8\linewidth}
        \centering
        \caption{}
        \includegraphics[width=0.9\linewidth]{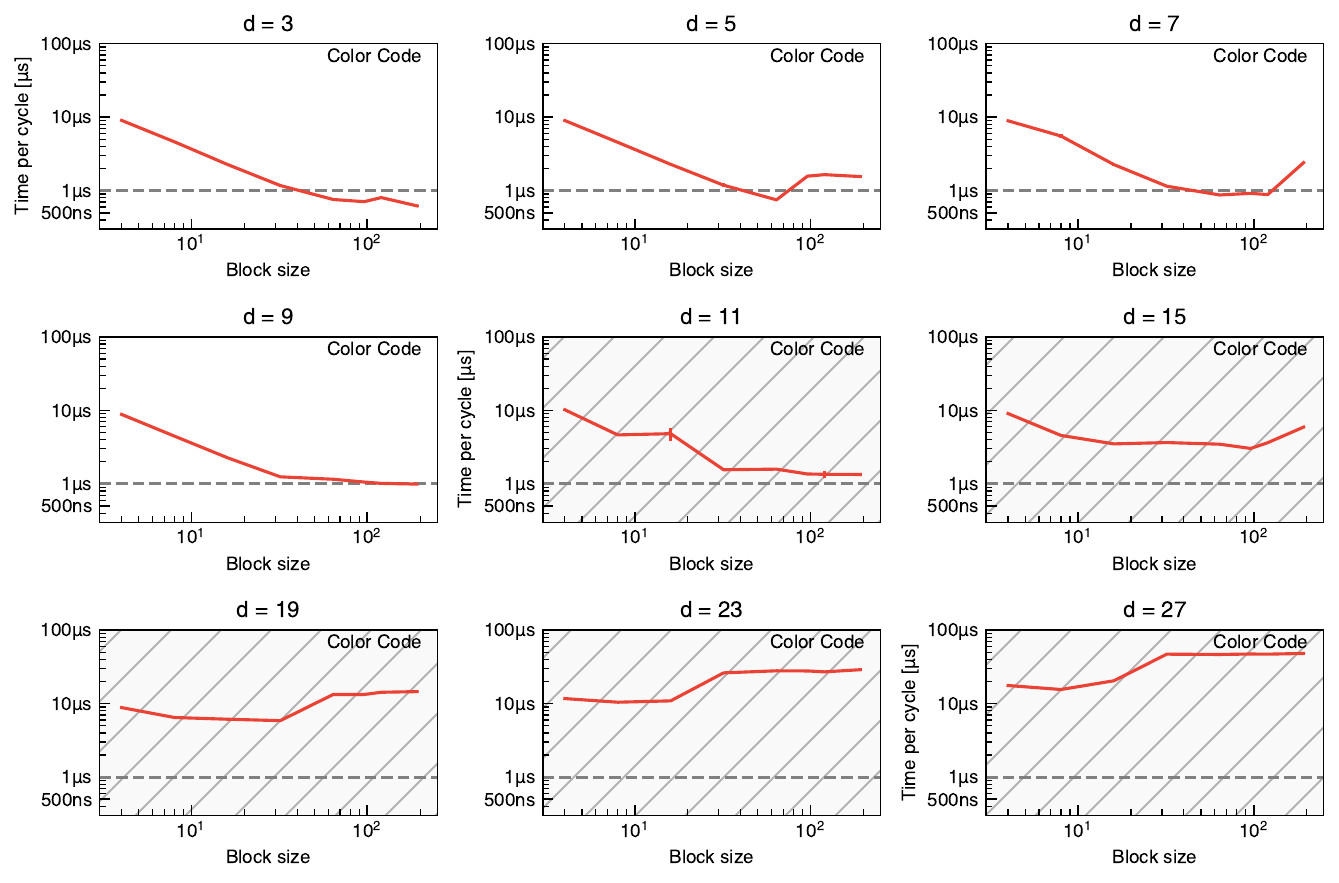}
        \label{fig:blocksize-cc}
    \end{subfigure}
    \caption{\textbf{Throughput measurements for different block sizes.}
    Average time per cycle for different block sizes for the AQ2-RT model for the surface code \textbf{(a)} and superdense colour code \textbf{(b)}. Shaded plots indicate distances for which the model was not trained.}
    \label{fig:throughputs-vs-blocksize}
\end{figure}

\begin{figure}[t]
\centering
    \begin{subfigure}[b]{0.45\linewidth}
    \centering
    \caption{}
    \includegraphics[scale=0.5]{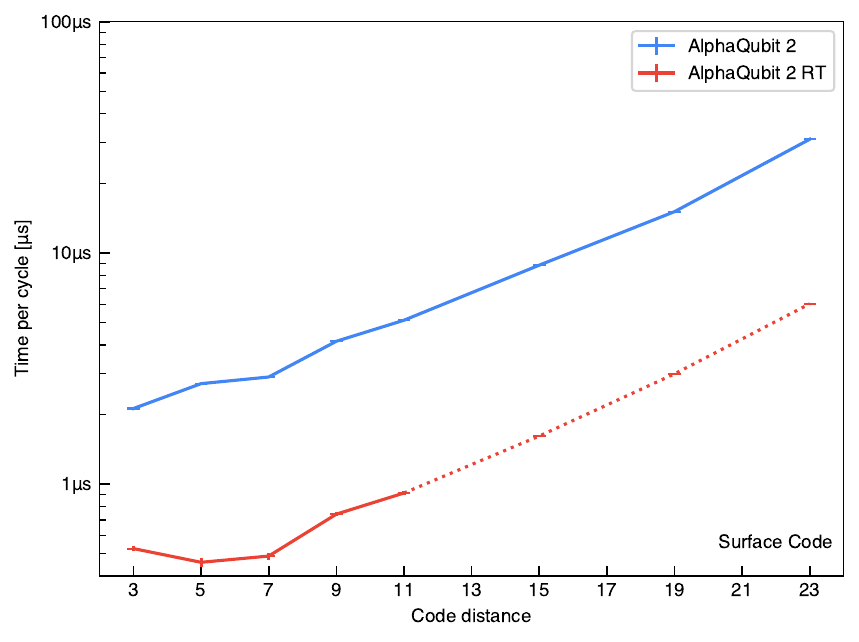}
    \label{fig:aqrt_vs_aq_sc}
    \end{subfigure}
    \hfill
    \begin{subfigure}[b]{0.45\linewidth}
    \centering
    \caption{}
    \includegraphics[scale=0.5]{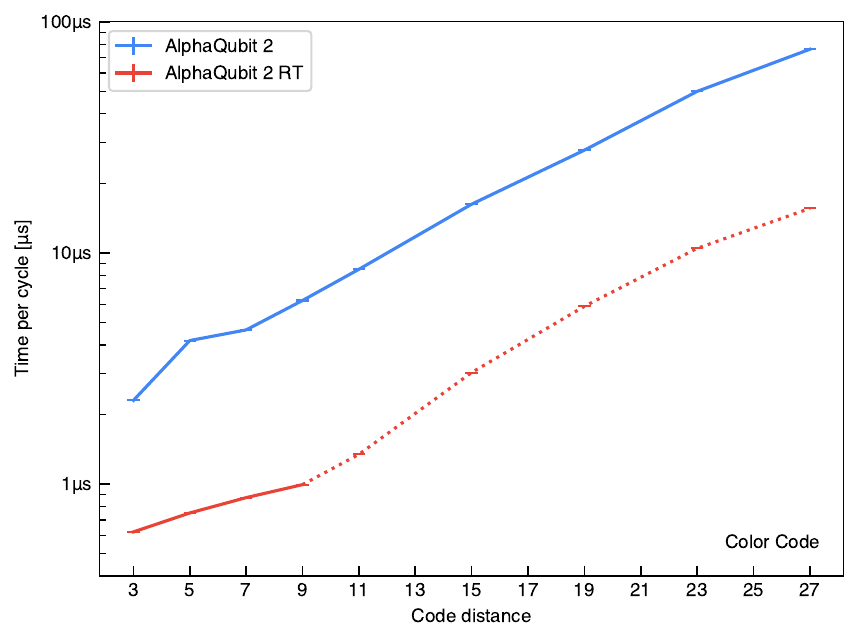}
    \label{fig:aqrt_vs_aq_cc}
    \end{subfigure}
\caption{\textbf{Throughput of AlphaQubit 2 (AQ2).} Average time per cycle for different code distances for AQ2-full and AQ2-RT for surface code (left) and superdense colour code (right). For each model and code distance, we use the block size (24, 48 or 96) resulting in the lowest time per cycle (\cref{fig:throughputs-vs-blocksize}). Error bars represent 1 standard deviation. Times for AQ2-RT are shown with dotted lines at code distances for which it has not been trained.}
\label{fig:throughputs_vs_full}
\end{figure}

\begin{figure}[t]
\centering
\includegraphics[scale=0.6]{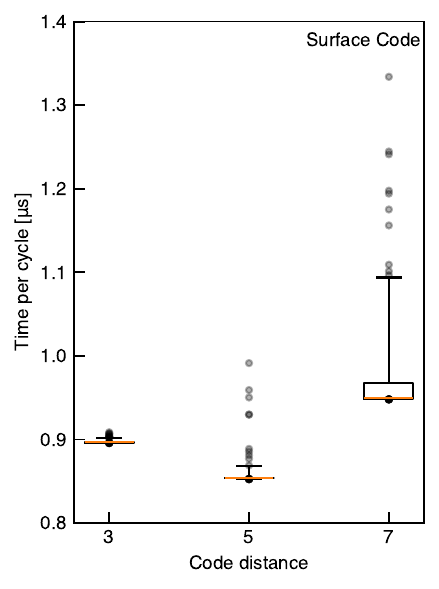}
\caption{\textbf{Willow real-time model decoding throughput.}
Box plot of throughputs for distance 3, 5 \& 7 models designed for real-time operation on the Willow data, compressing blocks of 5 adjacent frames. Each boxplot represents 1000 shots of 100800 cycles processing a block of 80 frames at once. Whiskers extend from 1st to 99th percentile. $896.6~ns \pm 1.1$ at distance 3; $854.3~ns \pm 7.1$ at 5; $965.2~ns \pm 3.4$ at 7. Measurements are taken on a Trillium TPU.}
\label{fig:y1_throughputs}
\end{figure}

\subsection{Latency}
Since the computational cost for every block of stabilizers and of the final readout network are both fixed, the throughput and latency of the decoder are constant. While we conduct our evaluations on simulated data, and are not measuring the latency involved in the measurement and communication hardware for a real quantum computer, we can measure the latency of the decoder itself as described in \ref{sec:measuring-speed}. 

We observe that the latencies in \cref{fig:throughputs}b and \cref{fig:throughputs-cc}b do not vary with the length of the experiment, showing that the decoder can keep up with a clock rate of $1~\upmu s$. While a small fraction of the shots show long latency, we attribute that to CPU scheduling issues that would be addressed by using a dedicated real-time operating system when running live experiments on a quantum computer.

\subsection{Achieving real-time throughput at distance 23}
AQ2 achieves higher throughputs than AQ1, due to temporal compression, parallelization across time of the spatial mixing transformer layers as well as a new hardware generation. We anticipate a number of techniques can be brought to bear to develop a faster network architecture and improve the throughput of a given architecture. 

The current streaming model operates with a \texttt{bfloat16} numerical representation and we expect speed-ups by adopting 8-bit or lower precision arithmetic, which has been the focus of much recent research, and is supported in the latest and forthcoming accelerators. We have also not yet applied accelerator kernel optimizations which we expect to bring significant speed-ups. Future generations of machine learning accelerators will also bring faster computation.

The current unoptimized AQ2-full achieves a throughput of one cycle every 30 microseconds with the surface code at distance 23. Although AQ2-RT is much faster when run at distance 23 ($6.02\pm 0.013\upmu s$ per cycle), it is not powerful enough to achieve useful error rates. We believe that we can train an intermediate scale model from the current architecture to decode distance 23 with similar accuracy to the current full model, but significantly faster. (For instance with a careful exploration of numbers of layers, embedding sizes and increasing the number of cycles combined together). 

Combining all these speed-ups may well bring us real-time decoding at distance 23 by the time that quantum computers reach this scale, but we believe that real-time decoding can be achieved by implementing a mature and optimized architecture in custom hardware (FPGA or ASIC).

\section{Baseline decoders}

As explained in the main paper, we demonstrate our decoder’s accuracy by comparison with baseline decoders for the corresponding codes.  For these baselines, we use a range of strong, recent decoders which are open source or to which we have access.  For the surface code we are able to use the open-source decoders PyMatching and Tesseract as well as an in-house decoder Libra and the real-time matching-based decoder. These provide us with a spectrum, from the fastest, but least accurate to the slowest, but most accurate: Real-time matching decoder, PyMatching, Correlated PyMatching, Libra, Tesseract.  As specified below, these decoders were run with suggested settings and on modern hardware, so we believe the timings are representative if not highly optimized.

For the colour code, we compared with Tesseract (\cref{fig:realtime}b). While Chromobius\cite{gidney2023chromobius} does support decoding the Bell-flagged colour code it does not use the flags themselves, and cannot decode the circuit that we have used. Chromobius performs better on the superdense colour code, so we use that for comparison, and note that for this code stabilizers in both bases can be read simultaneously so rounds are twice as fast as for the Bell-flagged colour code. 

\subsection{PyMatching}
We used PyMatching version 2.3.1 \cite{pymatching,higgott2025sparse}, which solves both (uncorrelated) minimum weight perfect matching (MWPM) and correlated MWPM using the blossom algorithm.
Timings were carried out on an Apple M1 Pro processor.
\subsection{Chromobius}
We used Chromobius\cite{gidney2023chromobius} version 1.1.0, also benchmarked on Apple M1 Pro processor.
\subsection{Tesseract}
We use the open-source Tesseract decoder\cite{Beni2025} \newline (version 0.1.1, commit-sha \texttt{21fc7b0e21a028e5a25ae7cc80fd845dcb154472} for surface code results and commit-sha \texttt{2905edc36ff5b5e983a16cc91c0c6b3f8e095d25} for colour code results). This was executed on Google Cloud CPUs with 64 threads using the recommended  “long-beam" configuration up to distance 11: 

\begin{table}
    \centering
    \begin{tabular}{ll}
    \toprule
Parameter     & Value \\
\midrule
pqlimit &1,000,000 \\
det\_beam & 20 \\
beam\_climbing & True \\
num\_det\_orders & 21 \\
det\_order & DetIndex \\
no\_revisit\_dets & True \\
det\_order\_seed & 0 \\\bottomrule
\end{tabular}
    \caption{\textbf{Parameters for Tesseract}}
    \label{tab:tesseractparams}
\end{table}

This configuration did not perform well with the colour code at distance 11 (needing a larger priority queue than previously needed for 120-cycle shots). Because of the computational resources required we did not evaluate beyond distance 11, and we extrapolate the accuracy based on a straight line fit of the distance 5 \& 7 log logical error per cycle to indicate the logical error rates that might be achievable without speed or memory constraints. 

For the surface code Tesseract and Libra performed similarly. Since Libra is much faster, we present results only for Libra.

\subsection{Libra}
Libra\cite{Libra} was run on a cloud-based Intel(R) Xeon(R) CPU @ 2.60GHz.

We use the “degeneracy” variant of Libra with an ensemble size of 100. The runtime of the Libra implementation we use is dominated by the computation of the complementary gap. We compute the complementary gap by modifying the detector error model to include the logical observable as a detector\cite{gidney2023yoked}. The resultant decoding problem breaks the assumptions required for sparse blossom to obtain almost-linear expected runtime \cite{higgott2025sparse}. Since the observable spans a boundary of the full 120-cycle experiment, this leads to the creation of a very large growing graph fill region. This graph fill region will grow to have very high blossom depth and saturate the $O(n^3)$ worst-case run time within the final augmentation stage of sparse blossom (where $n$ is the number of nodes in the DEM). This bottleneck could be removed by using a more efficient method for computing the complementary gap\cite{meister2024efficientsoftoutputdecoderssurface}.

\section{Statistics}

Throughout this article, we estimate confidence intervals in error rates using the Wilson score interval, as implemented by the statsmodels Python package v0.12.2.

\printbibliography[segment=\therefsegment,filter=notother]
\end{refsegment}

\end{document}